\documentclass[journal]{IEEEtran}
\usepackage[boxed]{algorithm}
\usepackage{algorithmic}
\usepackage{graphics}
\usepackage[dvips]{epsfig}
\usepackage[tight]{subfigure}
\usepackage{url}

\begin{document}
\title{Network Mapping by Replaying Hyperbolic Growth}
\author{Fragkiskos~Papadopoulos,
             Constantinos~Psomas,
             and~Dmitri~Krioukov%
\thanks{F. Papadopoulos and C. Psomas are with Cyprus University of Technology, Limassol,
Cyprus, e-mail: \{f.papadopoulos, c.psomas\}@cut.ac.cy.}%
\thanks{D. Krioukov is with the Cooperative Association for Internet Data Analysis (CAIDA), University of California, San Diego, USA, e-mail: dima@caida.org.}%
\thanks{A preliminary abstract version of this work appeared in~\cite{PaPsKr12}.}}

\maketitle

\begin{abstract}
Recent years have shown a promising progress in understanding geometric underpinnings behind the structure, function, and dynamics of many complex networks in nature and society. However these promises cannot be readily fulfilled and lead to important practical applications, without a simple, reliable, and fast network mapping method to infer the latent geometric coordinates of nodes in a real network.
Here we present \emph{HyperMap}, a simple method to map a given real network to its hyperbolic space. The method utilizes a recent geometric theory of complex networks modeled as random geometric graphs in hyperbolic spaces. The method replays the network's geometric growth, estimating at each time step the hyperbolic coordinates of new nodes in a growing network by maximizing the likelihood of the network snapshot in the model. We apply HyperMap to the AS Internet, and find that: 1)~the method produces meaningful results, identifying soft communities of ASs belonging to the same geographic region; 2)~the method has a remarkable predictive power: using the resulting map, we can predict missing links in the Internet with high precision, outperforming popular existing methods; and 3)~the resulting map is highly navigable, meaning that a vast majority of greedy geometric routing paths are successful and low-stretch. Even though the method is not without limitations, and is open for improvement, it occupies a unique attractive position in the space of trade-offs between simplicity, accuracy, and computational complexity.
\end{abstract}


\section{Introduction}
\label{sec:intro}

Our growing dependence on networks has inspired a burst of research activity in the field of network science. One focus of this research is to derive network models capable of explaining common structural characteristics of large real networks, such as the Internet, social networks, and many other complex networks~\cite{Dorogovtsev10-book,Lewis-book,DorMen-book03}.~\footnote{Here we use terms {\em complex networks\/} and {\em scale-free networks\/} interchangeably to mean real networks with distributions $P(k)$ of node degrees $k$ following power laws $P(k)\propto k^{-\gamma}$ (exponent $\gamma$ is usually between $2$ and $3$), and with strong clustering, i.e., with large numbers of triangular subgraphs~\cite{Dorogovtsev10-book}.} A particular goal is to understand how these characteristics affect the various processes that run on top of these networks, such as routing, information sharing, data distribution, searching, and epidemics~\cite{Dorogovtsev10-book,Lewis-book,BoSchu02-book}. Understanding the mechanisms that shape the structure and drive the evolution of real networks can also have important applications in designing more efficient recommender and collaborative filtering systems~\cite{MeEl11}, and for predicting missing and future links---an important problem in many disciplines~\cite{ClMo08,LuZhou11}.

Some fundamental connections between complex network topologies and hyperbolic geometry have been recently discovered in~\cite{KrPa10}. This work shows that random geometric graphs~\cite{Penrose03-book} in hyperbolic spaces are an adequate model for complex networks. The high-level explanation of this connection is that complex networks exhibit hierarchical, tree-like organization, while hyperbolic geometry is the geometry of trees~\cite{Gromov07-book}. Graphs representing complex networks appear then as discrete samples from the continuous world of hyperbolic geometry. The static approach in~\cite{KrPa10} has been extended to growing networks in~\cite{PaBoKr11}. This work shows that trade-offs between popularity and similarity shape the structure and dynamics of growing complex networks, and that these trade-offs in network dynamics give rise to hyperbolic geometry. The growing network model in~\cite{PaBoKr11} is nothing but a model of random geometric graphs growing in hyperbolic spaces. Synthetic graphs grown according to this simple model \emph{simultaneously} exhibit many common structural and dynamical characteristics of some real networks. Here we call the model in~\cite{PaBoKr11} the Popularity$\times$Similarity Optimization (PSO) model.

Given the ability of the PSO model to construct synthetic growing networks that resemble real networks across a wide range of structural and dynamical characteristics, can one reverse this synthesis, and given a real network, map (embed) the network into the hyperbolic plane, in a way congruent with the PSO model? Would the results of such mapping be meaningful? That is, can they be efficiently used in some applications, such as soft community detection, link prediction, or network navigation?

Here we give the affirmative answers to these questions. We first present a systematic framework to map a given complex network to its hyperbolic space, by replaying the network's geometric growth in accordance with the PSO model. The proposed network mapping method, called \emph{HyperMap}, is simple (cf.~Fig.~\ref{fig:the_method}) and supported by theoretical analysis (Section~\ref{sec:mapping_method}).  We apply HyperMap to the Autonomous Systems (AS) topology of the Internet to show that it produces meaningful results. It identifies soft communities of ASs belonging to the same geographic region. Given the Internet map constructed by HyperMap, we can predict missing links in the AS Internet with high precision by giving higher missing-link scores to disconnected node pairs located closer to each other. We show that this prediction yields better results than popular existing methods~\cite{LuZhou11}, especially for the links that are hard to predict. The AS Internet is known to be navigable~\cite{BoKrKc08,BoPa10}. Therefore the Internet map constructed by a good mapping method must be navigable as well. We show that this is indeed the case with HyperMap---greedy forwarding in the map can reach destinations with more than $90\%$ success probability and low stretch.

The rest of the paper is organized as follows. In Section~\ref{sec:preliminaries} we review the PSO model. In Section~\ref{sec:modified_pso_model} we introduce a modified version of this model, which is needed for an accurate replay of the hyperbolic growth of a given network. In Section~\ref{sec:mapping_method} we present the HyperMap method. In Section~\ref{sec:validation} we validate HyperMap on synthetic networks in the model. In Section~\ref{sec:real_nets} we apply the method to the real AS Internet, and show that it identifies soft communities of ASs belonging to the same country. In Section~\ref{sec:missing_link_prediction} we show that HyperMap predicts missing links in the Internet with high precision, and compare its performance against popular existing link-prediction methods. In Section~\ref{sec:navigation} we compute the navigability properties of the HyperMap-constructed map of the Internet. Finally, in Section~\ref{sec:discussion} we discuss open problems and conclude the paper.

\section{Preliminaries}
\label{sec:preliminaries}

In this section we review the PSO model~\cite{PaBoKr11}, limiting ourselves only to the basic details that we will need in the rest of the paper.

The \emph{basic} PSO model has four input parameters $m>0$, $\beta\in(0,1]$, $T \geq 0$, and $\zeta>0$. Parameter $m$ is the average number of existing nodes to which new nodes connect, defining the average node degree $\bar{k}=2m$ in the growing network. Parameter $\beta$ defines the exponent $\gamma=1+1/\beta\geq2$ of
the power-law degree distribution $P(k)\propto k^{-\gamma}$ in the network.~\footnote{Symbol ``$\propto$" means \emph{proportional to}, i.e., $f(t) \propto g(t)$ means $f(t)=c g(t)$, where $c$ is a constant, $0 < c < \infty$. Sometimes there are additive terms so that $f(t) \propto g(t)$ can also mean $f(t)=c g(t)+d$.  Symbol ``$\approx$" means \emph{approximately equal}. The approximations often become exact in the large graph size limit.} Temperature $T$ controls the average clustering~$\bar{c}$~\cite{Dorogovtsev10-book}
in the network, which is maximized at $T=0$, nearly linearly decreases to zero with $T\in[0,1)$, and is asymptotically zero if $T>1$. Parameter $\zeta=\sqrt{-K}$ where $K$ is the curvature of the hyperbolic plane. This parameter is dumb in the sense that it does not affect any properties of generated networks, so that it can be set to any value~\cite{PaBoKr11}, e.g., $\zeta=1$. However, we do not fix $\zeta$ to any value in our analysis below to make it more general. Having these parameters specified, the PSO model constructs a growing scale-free network up to $t > 0$ nodes according to the following {\bf PSO model definition}:
\begin{enumerate}
\item[(1)] initially the network is empty;
\item[(2)] coordinate assignment and update:
\begin{enumerate}
\item at time $i=1,2,\ldots,t$, new node $i$ is added to the hyperbolic plane at polar coordinates $(r_i, \theta_i)$, where radial coordinate $r_i=\frac{2}{\zeta}\ln{i}$, while the angular coordinate $\theta_i$ is sampled uniformly at random from $[0, 2\pi]$;
\item each existing node $j=1,2,\ldots,i-1$, moves increasing its radial coordinate according to
$r_j(i)=\beta r_j +(1-\beta)r_i$;
\end{enumerate}
\item[(3)] creation of edges: node $i$ connects to each existing node $j=1,2,\ldots,i-1$ with different probability $p_{ij} \equiv p(x_{ij})$ given by:
\begin{equation}
\label{eq:p_x_ji}
p(x_{ij})=\frac{1}{1+e^{\frac{\zeta}{2T}(x_{ij}-R_i)}}.
\end{equation}
\end{enumerate}
In the last expression, $x_{ij}$ is the hyperbolic distance between nodes $i$ and $j$~\cite{Bonahon09-book}:
\begin{eqnarray}
\label{eq:x_ji}
\nonumber x_{ij}=\frac{1}{\zeta}\mathrm{arccosh}\left(\cosh{\zeta r_i}\cosh{\zeta r_j}-\sinh{\zeta r_i} \sinh {\zeta r_j} \cos{\theta_{ij}}\right)\\
\nonumber \approx r_i+r_j+\frac{2}{\zeta}\ln{(\theta_{ij}/2)}, \quad\textnormal{where}~\theta_{ij}=\pi-|\pi-|\theta_i-\theta_j||,
\end{eqnarray}
while $R_i$ is derived from the condition that the expected number of nodes to which $i$ connects is indeed~$m$, yielding~\cite{PaBoKr11}:
\begin{equation}
\label{eq:R_i}
R_i=r_i-\frac{2}{\zeta}\ln\left[\frac{2T}{\sin{T\pi}}\frac{I_i}{m}\right],
\end{equation}
where $I_i=\frac{1}{1-\beta}(1-i^{-(1-\beta)})$.  Note that the appearance ``time" of a node is its \emph{order} of appearance in the network, i.e., the $i^{th}$ new node is said to appear at time $i$.

The radial coordinate of a node abstracts its popularity. The smaller the radial coordinate of a node, the more popular the node is, and the more likely it attracts new connections. The angular distance between two nodes abstracts their similarity. The smaller this distance, the more similar the two nodes are, and the more likely they are connected.  The hyperbolic distance $x_{ij}$ is then a single-metric representation of a combination of the two attractiveness attributes, radial popularity and angular similarity. The connection probability $p(x_{ij})$ is a decreasing function of $x_{ij}$, meaning that new connections take place by optimizing trade-offs between popularity and similarity~\cite{PaBoKr11}.

The connections between new nodes and existing nodes are called \emph{external} links. In many real networks however, certainly in the Internet, new links appear at a certain rate not only between new and old nodes, but also between old nodes only. The basic PSO model can be easily extended to account for such \emph{internal} links as well. This is done by the following additional step in the network construction process:
\begin{enumerate}
\item[(4)] at every time $i$, select a random pair of disconnected nodes $k, l < i$, and connect this pair with probability $p(x_{kl})=\frac{1}{1+e^{\frac{\zeta}{2T}(x_{kl}-R_i)}}$, repeating until $L > 0$ internal links are created.
\end{enumerate}
With internal links, the average node degree is $\bar{k}=2(m+L)$. Parameter $L$ is an additional parameter specifying the rate at which internal links appear, versus $m$, the external link rate. We call the PSO model that uses both external and internal links \emph{generalized} PSO model.

It has been shown that the generalized PSO model can reproduce not only the degree distribution and clustering of different real networks, but also several other important properties~\cite{PaBoKr11}.  Given the ability of the model to construct growing synthetic networks that resemble real networks,
in this paper we are interested in reversing the synthesis. Given a real network, such as the AS Internet, we want to map (embed) it into the hyperbolic plane, in a way congruent with the generalized PSO model. That is, we want to find the node radial and angular coordinates in the hyperbolic plane that maximize the probability that the given network is produced by the generalized PSO model. However, mapping a given network according to the generalized PSO model \emph{per se} is impossible for the following two reasons. The first is that there is no way to distinguish external from internal links given a single network topology snapshot. The second problem is that given a network snapshot, there is no way to learn the exact order of appearance (birth times) of nodes in the network, so we need a procedure that can estimate this order.

To tackle the first problem we introduce the E-PSO model in the next section. The E-PSO model is a model equivalent to the generalized PSO model, even though E-PSO uses \emph{external links only}. As a consequence of this equivalence, E-PSO can also simultaneously reproduce the same topological properties of the AS Internet as the generalized PSO. The second problem is addressed in Section~\ref{sec:mapping_method}, where we show that given the network topology, we can compute the maximum likelihood estimate (MLE) of the node appearance order. Using the MLE node appearance order, we can then map the AS Internet in a way congruent with the E-PSO model, treating all links in the topology \emph{as if} they were external.

\section{E-PSO: Growing Networks using External Links Only.}
\label{sec:modified_pso_model}

The E-PSO model is exactly the same as the basic PSO model described in the previous section, except that different nodes $i \leq t$ in E-PSO do not connect to the same expected number $m=\frac{\bar{k}}{2}$ of existing nodes $j < i$. Instead the expected number of connections that $i$ establishes is:
\begin{equation}
\label{eq:m_i}
\bar{m}_i(t)=m+\bar{L}_i(t),
\end{equation}
where parameter $m \leq \frac{\bar{k}}{2}$, while $\bar{L}_i(t)$ is the expected number of internal links between node $i$ and existing nodes $j  <  i$ by time $t$, in the generalized PSO model.

To compute this number, we start with the probability that a pair of existing nodes $i, j$ establishes an internal link at time $l$ in the generalized PSO (\cite{PaBoKr11}, Supplementary Information, Section VIII):
\begin{equation}
\label{eq:attraction_prob}
\Pi(i, j, l)=2L\frac{e^{-\frac{\zeta}{2}(r_i(l)+r_j(l))}}{\int_{1}^{l} \int_{1}^{l} e^{-\frac{\zeta}{2}(r_i(l)+r_j(l))}didj}=2L\frac{l^{2\beta-2}(ij)^{-\beta}}{I_l^2},
\end{equation}
where $r_i(l)$, $i\leq l$, is the radial coordinate of node $i$ at time $l$, and $I_l=\frac{1}{1-\beta}(1-l^{-(1-\beta)})$.
Using Equation~(\ref{eq:attraction_prob}) we can compute the probability that $i$ and $j$ are connected by an internal link by time $t$, if $j < i$:
\begin{eqnarray}
\nonumber \tilde\Pi(i,j,t)&=&\int_{i}^{t}\Pi(i, j, l)dl\\
\nonumber &\approx& \frac{2L (1-\beta)^2}{(1-t^{-(1-\beta)})^2(2\beta-1)}(ij)^{-\beta}(t^{2\beta-1}-i^{2\beta-1}),
\end{eqnarray}
where the  approximation uses the fact that for large $l ,t$, $I_l\approx I_t$.
Therefore, the expected number of internal links between node $i$ and all previous nodes $j  <  i$ by time $t$, is:
\begin{eqnarray}\label{eq:L_i}
\nonumber \bar{L}_i(t)&=&\int_{1}^{i}\tilde\Pi(i,j,t)dj\approx\frac{2L (1-\beta)}{(1-t^{-(1-\beta)})^2(2\beta-1)}\\
&\times& \left[\left(\frac{t}{i}\right)^{2\beta-1}-1\right]\left[1-i^{-(1-\beta)}\right].
\end{eqnarray}
Limits $\beta \to 1$ and $\beta \to 0.5$ in the above relation are: $\bar{L}_i(t)  \to 2L\frac{(t-i)\ln{i}}{i(\ln{t})^2}$  if $\beta \to 1$, and $\bar{L}_i(t) \to L \frac{1-i^{-0.5}}{(1-t^{-0.5})^2}\ln{(\frac{t}{i})}$ if $\beta \to 0.5$. As in the PSO models, we can show, see Appendix, that in E-PSO the expected degree of node $i$ by time $t$, $\bar{k}_i(t)$, satisfies:
\begin{equation}
\label{eq:k_i_t}
\bar{k}_i(t) \propto \left(\frac{i}{t}\right)^{-\beta},
\end{equation}
which means that the degree distribution in E-PSO is also a power law, $P(k) \propto k^{-\gamma}$, with $\gamma=1+\frac{1}{\beta}$. Further, $\bar{k} \approx 2(m+L)$. We note that if $L=0$ then $\bar{L}_i(t)=0, \forall i$, and the E-PSO model degenerates to the basic PSO model.

Summarizing, the E-PSO has five input parameters $m, L, \beta, T, \zeta$, and to construct a network up to $t$ nodes, one follows exactly the same procedure as in the basic PSO, except that $R_i$ in Equation~(\ref{eq:R_i}) is adjusted to:
\begin{equation}
\label{eq:R_i_new}
R_i=r_i-\frac{2}{\zeta}\ln\left[\frac{2T}{\sin{T\pi}}\frac{I_i}{\bar{m}_i(t)}\right],
\end{equation}
with $\bar{m}_i(t)$ in Equation~(\ref{eq:m_i}) and $\bar{L}_i(t)$ in Equation~(\ref{eq:L_i}).

\begin{figure*}
\centerline{
\subfigure[Degree distribution $P(k)$.]{\includegraphics [width=2.2in, height=1.2in]{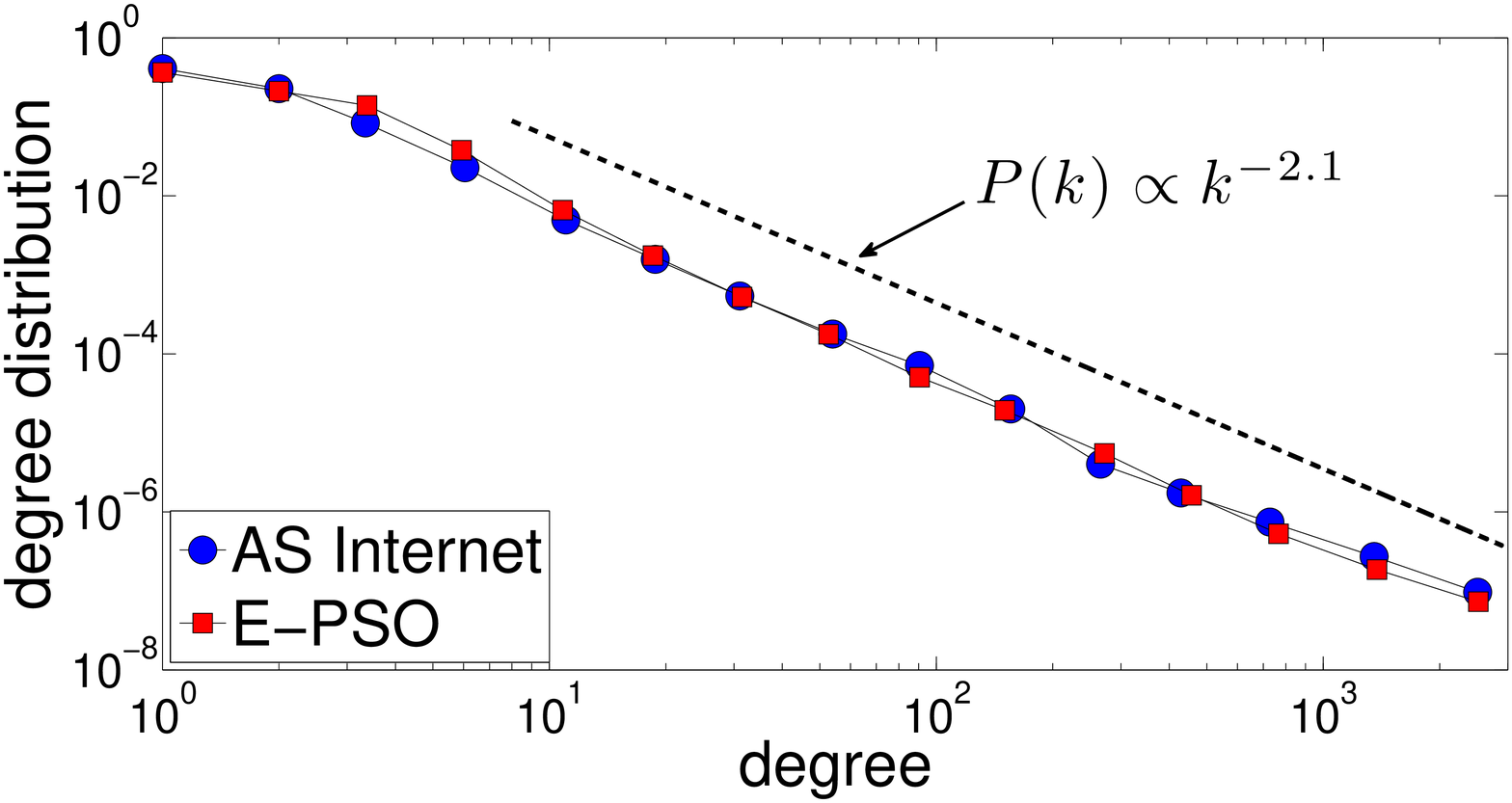}}
\subfigure[Average clustering $\bar{c}(k)$.]{\includegraphics [width=2.2in, height=1.2in]{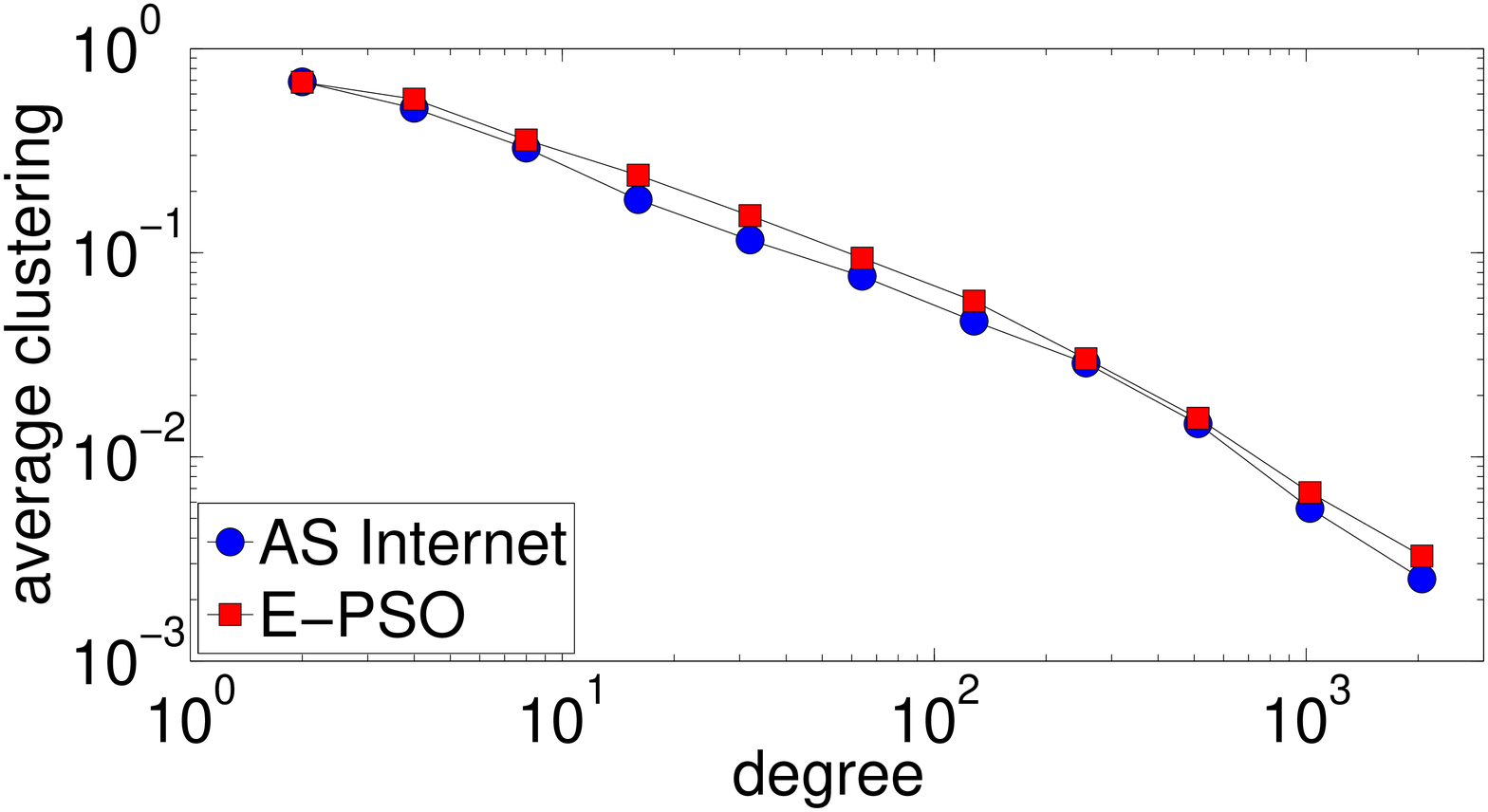}}
\subfigure[Average neighbor degree $\bar{k}_{nn}(k)$.]{\includegraphics [width=2.2in, height=1.2in]{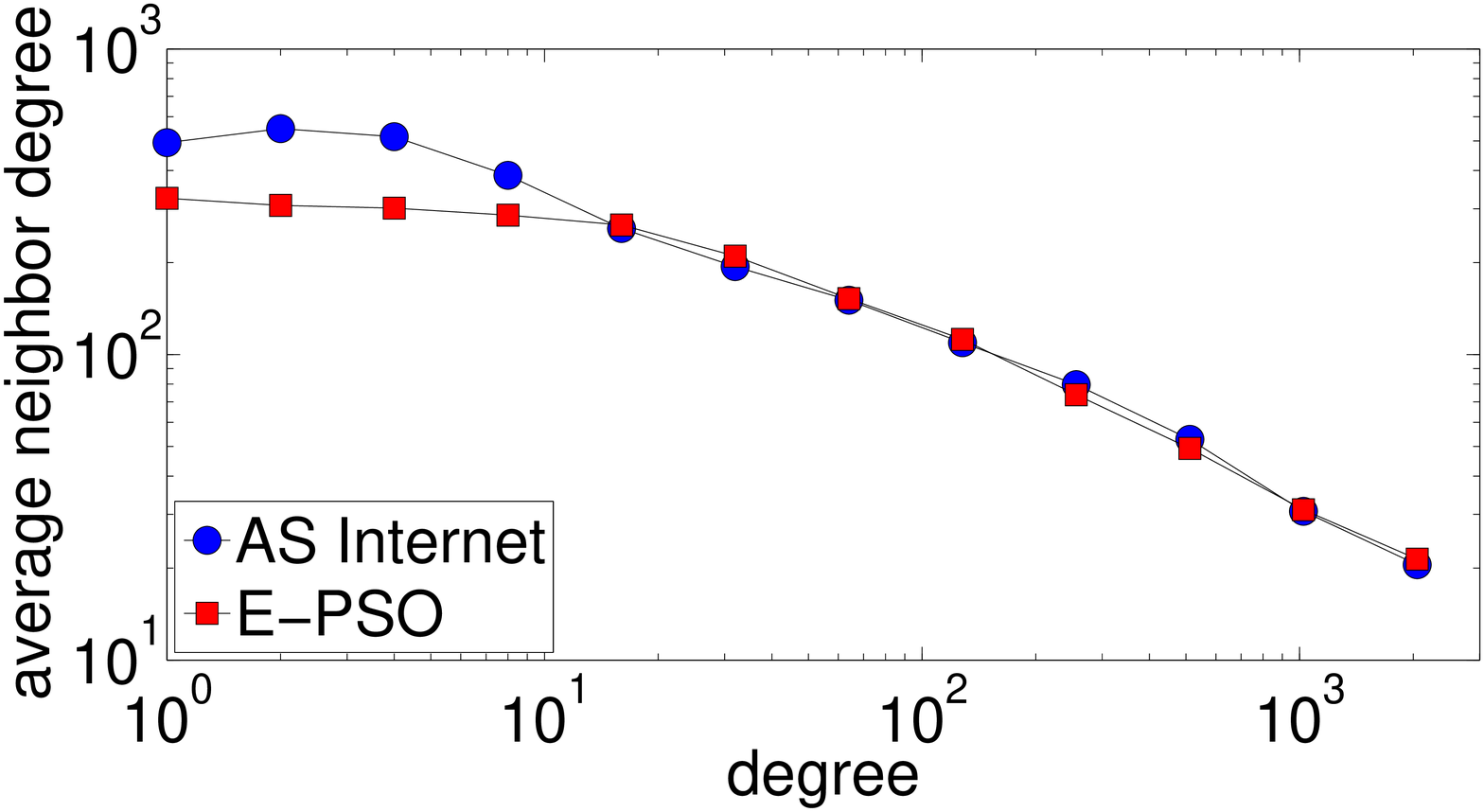}}}
\centerline{
\subfigure[Distance distribution $d(l)$.]{\includegraphics [width=2.2in, height=1.2in]{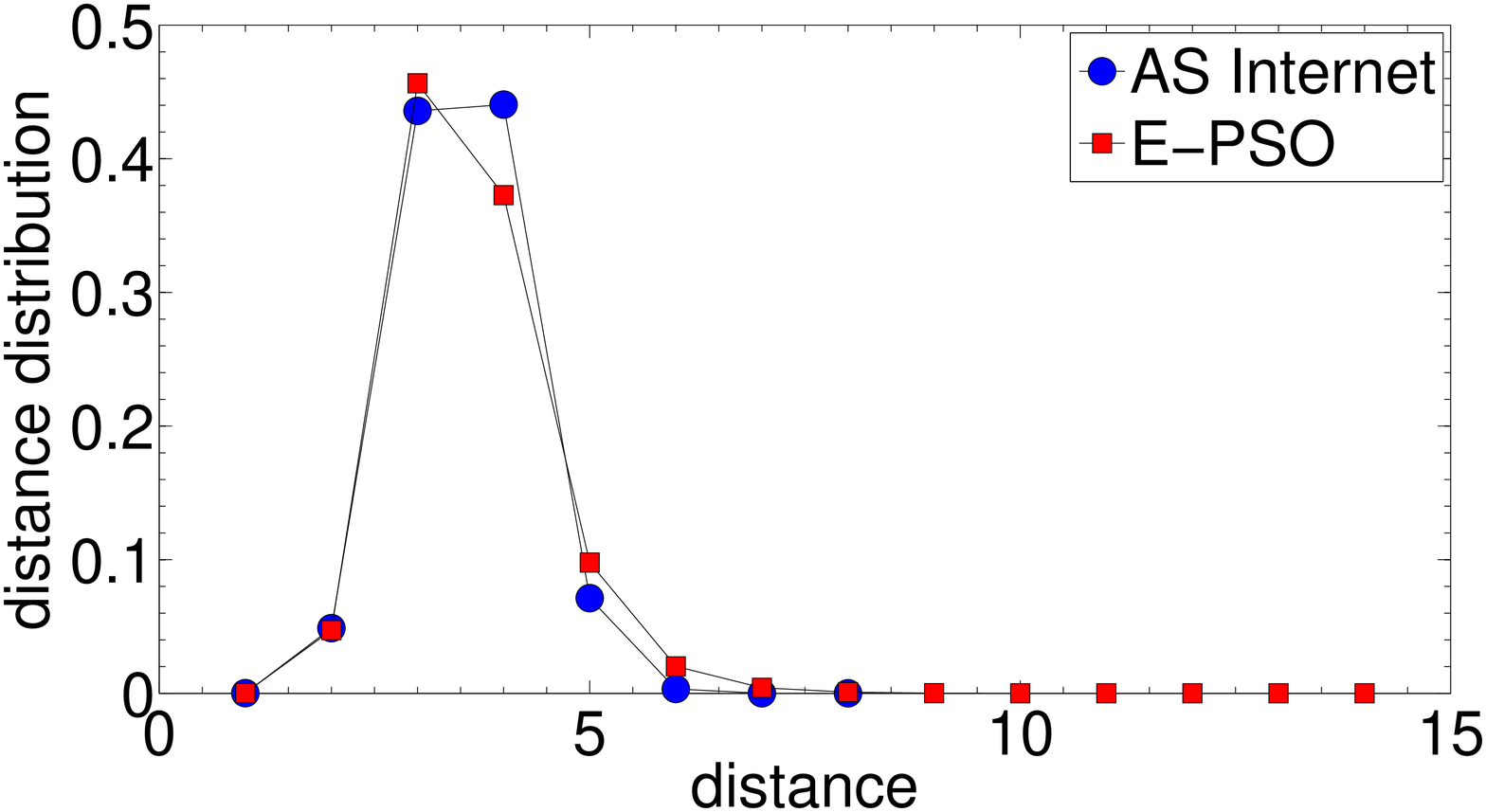}}
\subfigure[Average betweenness $\bar{B}(k)$.]{\includegraphics [width=2.2in, height=1.23in]{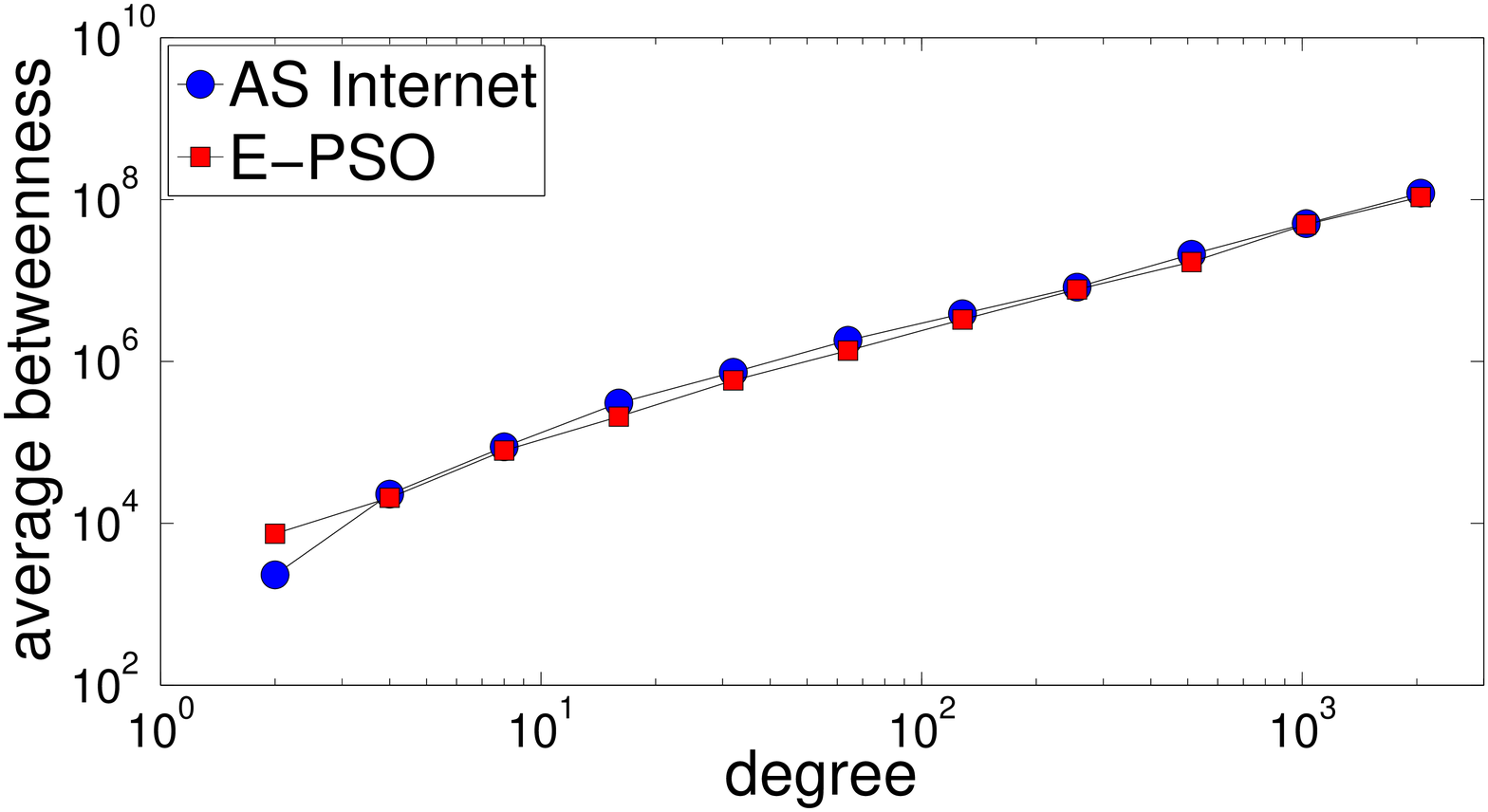}}
\subfigure[$\tilde{m}_i(t)$ vs.\ node appearance time~$i$.]{\includegraphics [width=2.2in, height=1.18in]{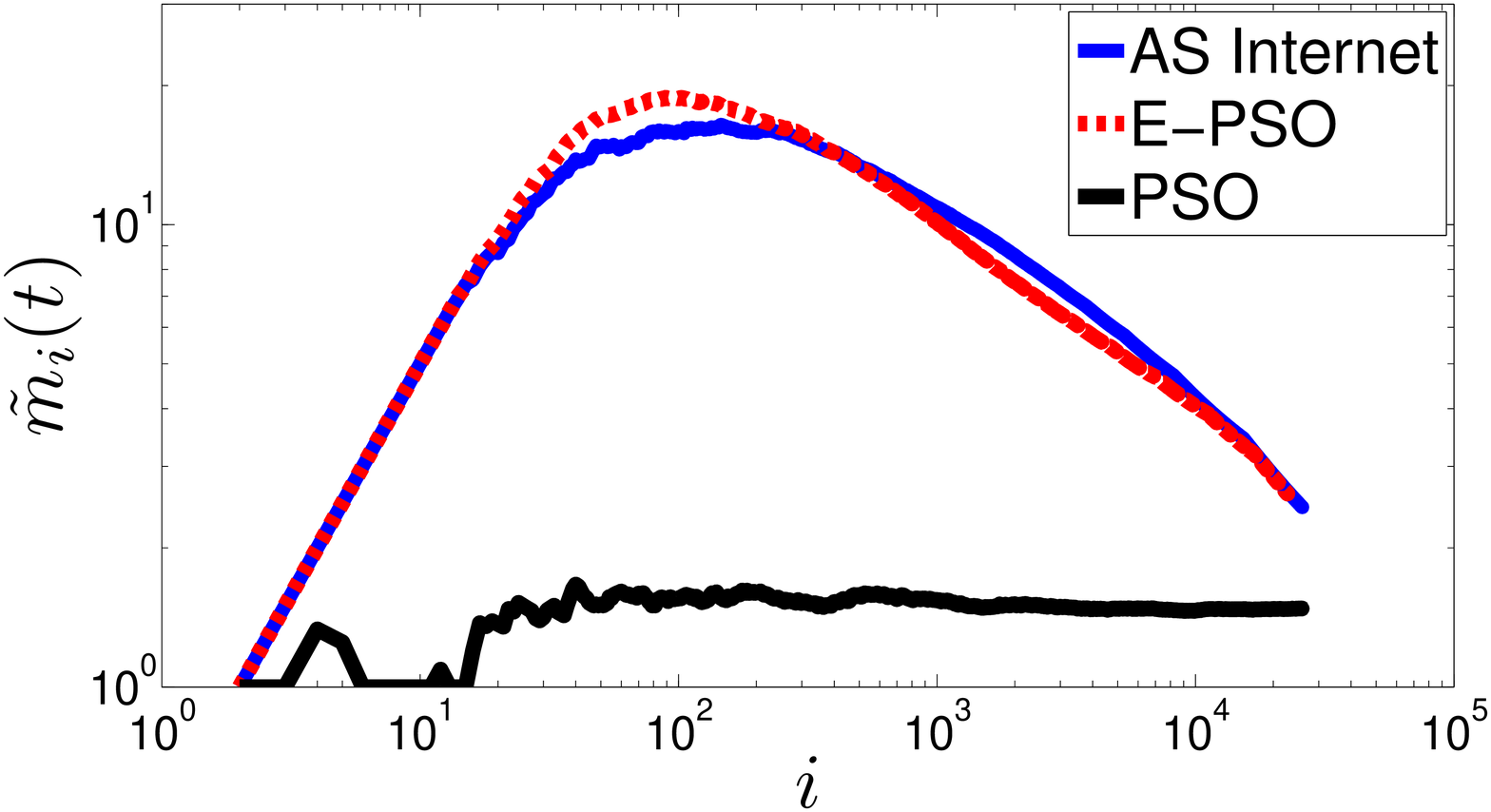}}}
\caption{Properties of the AS Internet vs.\ simulated networks grown according to the E-PSO model.}
\label{fig:inet_vs_sim}
\end{figure*}

\textbf{Validation.} Figure~\ref{fig:inet_vs_sim} compares several important properties of simulated networks growing according to E-PSO to the properties of the AS Internet topology~\cite{ClHy09} of December 2009, which is available at~\cite{as_topo_data}. The topology consists of $t=25910$ nodes (ASs), and has a power law degree distribution with exponent $\gamma=2.1$, average node degree $\bar{k} \approx 5$ and average clustering $\bar{c}=0.61$. The connections in the topology are not physical but logical, representing AS relationships~\cite{as_topo_data}. Using the real data of the twelve-year (1998-2010) evolution of the AS Internet from~\cite{inet_evol_paper} we find that the average initial number of connections of an AS is $m \approx 1.5$, which means that $L = \frac{\bar{k}-2m}{2} \approx 1$.~\footnote{The data of~\cite{inet_evol_paper} are spaced by three-month intervals. We take as initial number of connections of an AS the number of connections the AS has when it is first seen in the data.} The simulated E-PSO network is grown up to the same number of nodes $t$ as in the real AS Internet and has the same $m, L, \gamma$ and $\bar{c}$. To yield $\bar{c}=0.61$, we set $T=0.45$.

Figure~\ref{fig:inet_vs_sim} considers the following properties, as in~\cite{PaBoKr11}: (a) the degree distribution $P(k)$; (b) the average clustering $\bar{c}(k)$ of $k$-degree nodes; (c) the average degree of neighbors $\bar{k}_{nn}(k)$ of $k$-degree nodes; (d) the distance distribution $d(l)$, i.e., the distribution of hop lengths $l$ of shortest paths between nodes in the network; and (e) the average node betweenness $\bar{B}(k)$ of $k$-degree nodes, which is the average number of shortest paths passing through a $k$-degree node, normalized by the maximum possible number of such paths. Properties (a-c) are local statistics reflecting properties of individual nodes and their one-hop
neighborhoods, as opposed to global properties (d-e). From the figure, we observe a remarkable match between the AS Internet and the simulated E-PSO network across all five properties. We emphasize that to accurately match all these properties in \cite{PaBoKr11} the generalized PSO model had to be used, which uses both external and internal links (see Fig.~S11 in~\cite{PaBoKr11}). By contrast, here we show that we can accurately match the same properties with E-PSO that uses external links only (Figures~\ref{fig:inet_vs_sim}(a)-(e)).~\footnote{Proving that the generalized PSO and E-PSO models can reproduce the same graph properties is beyond the scope of this paper. The proof consists of showing that the generalized PSO satisfies Equation~(\ref{eq:global_prob}).}

Further, for each node $i=2,3, \ldots, t$ in the simulated E-PSO network we also measure the number of links $m_i(t)$ to old nodes $j < i$, and compute its moving average $\tilde{m}_i(t)=\frac{1}{i-1}\sum_{j=2}^{i}m_j(t)$. We also compute $\tilde{m}_i(t)$ for the AS Internet after assuming that nodes with higher degrees appear earlier. (See the next section for the reason behind this assumption.) We use $\tilde{m}_i(t)$ as a summary statistic to validate Equations~(\ref{eq:m_i},\ref{eq:L_i}) in the AS Internet, by comparing its value to that in the simulated network. The results are shown in Figure~\ref{fig:inet_vs_sim}(f), where we again see a remarkable match between the AS Internet and the E-PSO network. The figure also reports the results for a simulated network grown according to the generalized PSO model with the same parameters. In this case each new node, upon its appearance, connects to the same average number of existing nodes $m=1.5$, i.e., $\tilde{m}_i(t) \approx m=1.5$.

\begin{figure}[!ht]
\centerline{\includegraphics [width=2.8in, height=1.7in]{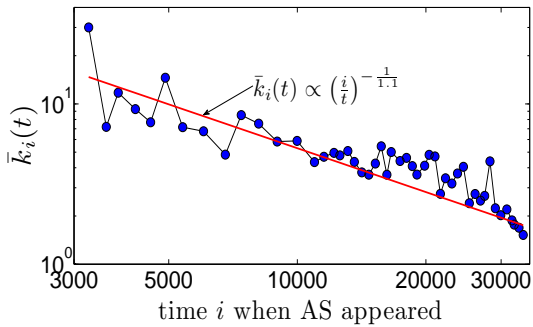}}
\caption{The average degree of ASs as a function of their birth times.}
\label{fig:inet_evol}
\end{figure}

Finally, in Figure~\ref{fig:inet_evol} we use the data from~\cite{inet_evol_paper} to validate that Equation~(\ref{eq:k_i_t}) indeed describes the trend in the evolution of the average degree of an AS in the Internet as a function of the time the AS appeared. To draw Figure~\ref{fig:inet_evol} we first found from the data in~\cite{inet_evol_paper} the time $i$  (number of nodes present in the network), when each AS first appeared in the data. Then, for all ASs that appeared at time $i$ and that are still present at the end of the measurement period where $t=33796$ nodes, we calculated their average degree $\bar{k}_i(t)$ as a function of their birth time~$i$. In the theoretical formula in Eq.~(\ref{eq:k_i_t}) we use the $\gamma$ of the AS Internet, i.e., $\beta=\frac{1}{\gamma-1}=\frac{1}{1.1}$.

Given the ability of the E-PSO model to construct growing synthetic networks that resemble real networks, such as the AS Internet, we next show that it is possible to reverse the synthesis, and given the AS Internet to map (embed) it into the hyperbolic plane, in a way congruent with the E-PSO model.

\section{HyperMap: Network Mapping by Replaying Hyperbolic Growth}
\label{sec:mapping_method}

In this section we present HyperMap, a method that computes radial and angular coordinates $\{r_i(t), \theta_i\}$ for all nodes $i=1,\ldots,t$ in a given network of size $t$ with adjacency matrix $\alpha_{ij}$---$\alpha_{ij}=\alpha_{ji}=1$ if there is a link between nodes $i$ and $j$, and $\alpha_{ij}=\alpha_{ji}=0$ otherwise.~\footnote{In this paper, notation ``\{~\}" denotes a set. For example, $\{r_i(t), \theta_i\}=r_1(t), \theta_1, r_2(t), \theta_2,\ldots, r_t(t), \theta_t$.}  Contrary to the previous sections and unless noted otherwise, the numbering of nodes in this section is arbitrary and unrelated to the order of appearance of nodes in the network. HyperMap is based on \emph{Maximum Likelihood Estimation}: it finds the node coordinates in the network by maximizing the probability, or likelihood, that the network is produced by the E-PSO model. Therefore the better the E-PSO model describes a given network, the better the quality of the mapping. We first give the necessary definitions and derive the likelihood that HyperMap maximizes.

\subsection{Definitions and Likelihood}

\subsubsection{\underline{Joint probability density of node coordinates}}

Recall that in E-PSO the node angular coordinates are random numbers sampled from the uniform distribution on $[0, 2\pi]$, i.e., their probability density is $\rho(\theta)=\frac{1}{2\pi}$.  In the Appendix we also derive the probability density of the node radial coordinate at time $t$:
\begin{equation}
\label{eq:f_r_i_t}
f_t(r)=\frac{\zeta}{2\beta}e^{\frac{\zeta}{2\beta}(r-r_t)}=\frac{\zeta(\gamma-1)}{2}e^{\frac{\zeta(\gamma-1)}{2}(r-r_t)},
\end{equation}
where $r_t=\frac{2}{\zeta}\ln{t}$. We note that the node coordinates in E-PSO are independent variables. Therefore, given $\rho(\theta)$ and $f_t(r)$, the joint probability that the node coordinates take the values $\{r_i(t), \theta_i\}$ is:
\begin{equation}
\label{eq:joint_density}
\textnormal{Prob}(\{r_i(t), \theta_i\} | \gamma, \zeta)=\frac{1}{(2\pi)^t} \prod_{i=1}^{t} f_t(r_i(t)).
\end{equation}

\subsubsection{\underline{Global and local connection probabilities}}

Consider a network that has grown up to $t$ nodes according to E-PSO. The \emph{global} connection probability $\tilde{p}(x(t))$ is the probability that  two random nodes at hyperbolic distance $x(t)$ are connected. In the Appendix we show that:
\begin{eqnarray}
\label{eq:global_prob}
\nonumber \tilde{p}(x(t))&= &\frac{1}{t-i_{min}+1} \sum_{i=i_{min}}^{t}\frac{1}{1+e^{\frac{\zeta}{2T}(x(t)-R_t+\Delta_i(t))}}\\
&\approx&\frac{1}{1+e^{\frac{\zeta}{2T}(x(t)-R_t)}},
\end{eqnarray}
where $i_{min}=\max(2, \lceil te^{-\frac{\zeta x(t)}{4(1-\beta)}}\rceil)$, $R_t$ given by Equation~(\ref{eq:R_i_new}), $\Delta_i(t)=\frac{2}{\zeta}\ln{\left[\left(\frac{t}{i}\right)^{2\beta-1}\frac{m I_i}{\bar{m}_i(t) I_t}\right]}$, $\bar{m}_i(t)$ given by Equation~(\ref{eq:m_i}), and $I_i=\frac{1}{1-\beta}(1-i^{-(1-\beta)})$. We call $\tilde{p}(x(t))$ \emph{global} because it is computed over all node pairs whose hyperbolic distance at time $t$ is $x(t)$. On the other hand, $p(x_{ij})$ in Equation~(\ref{eq:p_x_ji}) is called \emph{local} as it refers to the specific pair of nodes $i, j$, whose hyperbolic distance when $i$ appears is $x_{ij}$.

\subsubsection{\underline{Global likelihood}}

Consider a network that has grown up to $t$ nodes according to E-PSO with parameters $m, L, \gamma, T, \zeta$, and let $\alpha_{ij}$ be the resulting network adjacency matrix. We denote by $\mathcal L_1 \equiv \mathcal L(\{r_i(t), \theta_i\} | \alpha_{ij}, m, L, \gamma, T, \zeta)$ the likelihood that the node coordinates take the particular values $\{r_i(t), \theta_i\}$ given $\alpha_{ij}$ and $m, L, \gamma, T, \zeta$. Using Bayes' rule we can rewrite $\mathcal L_1$ as:
\begin{equation}
\label{eq:likelihood_1}
\mathcal L_1=\frac{\textnormal{Prob}(\{r_i(t), \theta_i\} | \gamma, \zeta) \mathcal L_2}{\mathcal L_3},
\end{equation}
where $\textnormal{Prob}(\{r_i(t), \theta_i\} | \gamma, \zeta)$ is given by Equation~(\ref{eq:joint_density}); $\mathcal L_2 \equiv \mathcal L(\alpha_{ij} | \{r_i(t), \theta_i\}, m, L, \gamma, T, \zeta)$ is the likelihood  to have the network with adjacency matrix $\alpha_{ij}$ if the node coordinates have the values $\{r_i(t), \theta_i\}$ and the parameters are $m, L, \gamma, T, \zeta$; and $\mathcal L_3 \equiv \mathcal L(\alpha_{ij} | m, L, \gamma, T, \zeta)$, independent of  $\{r_i(t), \theta_i\}$, is the probability that the E-PSO model with the given parameters generates the network with $\alpha_{ij}$.
We can compute $\mathcal L_2$ using Equation~(\ref{eq:global_prob}):
 \begin{equation}
\label{eq:likelihood_2}
\mathcal L_2=\prod_{1 \leq j < i \leq t} \tilde{p}(x_{ij}(t))^{\alpha_{ij}}\left[1-\tilde{p}(x_{ij}(t))\right]^{1-\alpha_{ij}},
\end{equation}
where the product goes over all node pairs $i, j$ in the network, and $x_{ij}(t)$ is the hyperbolic distance between pair $i, j$.~\footnote{For example, in a network with $t=3$ nodes, $1,2,3$, where only nodes $1$--$2$ and $1$--$3$ are connected, i.e., $\alpha_{12}=\alpha_{13}=1$, $\alpha_{23}=0$, $\mathcal L_2$ would be $\mathcal L_2=\tilde{p}(x_{12}(t))\tilde{p}(x_{13}(t))[1-\tilde{p}(x_{23}(t))]$.} We note that according to the model definition, all edges $\alpha_{ij}=1$ and non-edges $\alpha_{i'j'}=0$ are independent, and exist or non-exist with different probabilities $\tilde{p}(x_{ij}(t))$ and $1-\tilde{p}(x_{i'j'}(t))$, which depend on the hyperbolic distance between nodes. Since all the (non-)edges are independent, we can multiply the probabilities in Equations~(\ref{eq:likelihood_2},\ref{eq:local_likelihood_2}).

\subsubsection{\underline{Local likelihood}}

In contrast to the global likelihood that corresponds to the whole network at the final time $t$, the local likelihood is defined on a per-node basis as the network grows. Specifically, consider new node $i \leq t$ in a network that grows according to E-PSO, where nodes are now numbered according to the order they appear. When node $i$ appears, its radial coordinate is $r_i=\frac{2}{\zeta}\ln{i}$. We denote by $\mathcal L_1^{i} \equiv \mathcal L(\theta_i |r_i, \{r_j(i), \theta_j\}, \alpha_{ij}, m, L, \gamma, T, \zeta)_{j<i}$ the likelihood that $i$'s angular coordinate takes value $\theta_i$, given its $r_i$, the coordinates of the old nodes $\{r_j(i), \theta_j\}$, $j < i$, $i$'s connections to the old nodes $j<i$ in $\alpha_{ij}$, and the parameters $m, L, \gamma, T, \zeta$. Using Bayes' rule, we have:
\begin{equation}
\label{eq:local_likelihood_1}
\mathcal L_1^{i}=\frac{1}{2\pi}\frac{\mathcal L_2^{i}}{\mathcal L_3^{i}},
\end{equation}
where $\mathcal L_2^{i} \equiv \mathcal L(\alpha_{ij} | r_i, \theta_i, \{r_j(i), \theta_j\}, m, L, \gamma, T, \zeta)_{j<i}$ is the likelihood to have the connections $\alpha_{ij}$, $j <i$, if the angular coordinate of node $i$ has value $\theta_i$, conditioned on its radial coordinate, the coordinates of the old nodes, and the network parameters. Likelihood $\mathcal L_3^{i} \equiv \mathcal L(\alpha_{ij} | r_i, \{r_j(i), \theta_j\}, m, L, \gamma, T, \zeta)_{j<i}$, independent of $\theta_i$, is the probability that $i$ has the connections specified by $\alpha_{ij}$, $j<i$, conditioned as shown by notation. We can compute $\mathcal L_2^{i}$ using Equation~(\ref{eq:p_x_ji}):
\begin{equation}
\label{eq:local_likelihood_2}
\mathcal L_2^{i}=\prod_{1 \leq j < i} p(x_{ij})^{\alpha_{ij}}\left[1-p(x_{ij})\right]^{1-\alpha_{ij}}.
\end{equation}
The product goes over all the old nodes $j < i$.

\subsection{Likelihood Maximization}

We are looking for the values $\{r_i(t)^{*}, \theta_i^{*}\}$ that maximize the global likelihood $\mathcal L_1$ in Equation~(\ref{eq:likelihood_1}), or equivalently,
its logarithm:
\begin{eqnarray}
\label{eq:log_likelihood_1}
\nonumber\ln{\mathcal L_1}&=&C+\frac{\zeta}{2\beta}\sum_{i=1}^t r_i(t)+\sum_{i=1}^{t-1}\sum_{j=i+1}^{t}\alpha_{ij}\ln{\tilde{p}(x_{ij}(t))}\\
&+&\sum_{i=1}^{t-1}\sum_{j=i+1}^{t}(1-\alpha_{ij})\ln{[1-\tilde{p}(x_{ij}(t))]},
\end{eqnarray}
where $C$ is a constant independent of $\{r_i(t), \theta_i\}$.  Unfortunately, the maximization of Equation~(\ref{eq:log_likelihood_1}) can be performed analytically with respect to $r_i(t)$ only, but not with respect to $\theta_i$. Another problem is that even though there are plenty of methods to numerically find maximum-likelihood solutions, e.g., Markov Chain Monte Carlo (MCMC) methods such as the Metropolis-Hastings algorithm~\cite{Newman99-book},
these methods do not provide any reasonable performance guarantees. They have \emph{exponential} worst-case running times, and require significant manual intervention and guidance to lead to any reasonable results in a reasonable amount of compute time~\cite{BoPa10}. We do not follow this approach here.

Instead we first use Equation~(\ref{eq:log_likelihood_1}) to analytically find the maximum likelihood estimate (MLE) of the sequence according to which nodes appeared in a given network. From this sequence we then compute $\{r_i(j)^{*}\}$, $\forall~j\leq t$, and replay the growth of the network according to the E-PSO model, finding for each new node $i$ its angle ${\theta_i^{*}}$ that maximizes the local likelihood $\mathcal L_1^{i}$ in Equation~(\ref{eq:local_likelihood_1}), or equivalently, $\mathcal L_2^{i}$ in Equation~(\ref{eq:local_likelihood_2}). Maximizing the local likelihood at each time $i \leq t$ is equivalent to maximizing the global likelihood at the final time $t$. This approach leads to HyperMap, which performs remarkably well in finding $\{r_i(t)^{*}, \theta_i^{*}\}$ and has a \emph{guaranteed} running time. We proceed with the MLE of the node appearance times.

\subsection{MLE of node appearance times}

The derivative of Equation~(\ref{eq:log_likelihood_1}) with respect to $r_i(t)$ gives:
\begin{equation}
\label{eq:d_log_likelihood_1}
{\frac{\partial \nonumber\ln{\mathcal L_1}}{\partial r_i(t)}}=\frac{\zeta}{2\beta}-\frac{\zeta}{2T}\left(\sum_{j=1,j\ne i}^{t}\alpha_{ij}-\sum_{j=1,j\ne i}^{t}\tilde{p}(x_{ij}(t))\right).
\end{equation}
The first sum within the parenthesis is the actual degree of node $i$, $k_i$, while the second sum is its expected degree~$\tilde{\bar{k}}_i(t)$.
The likelihood is maximized when ${\frac{\partial \nonumber\ln{\mathcal L_1}}{\partial r_i(t)}}=0$, i.e., when
\begin{equation}
\label{eq:ml_k_i_t}
\tilde{\bar{k}}_i(t)=k_i-\frac{T}{\beta}.
\end{equation}
Expected degree $\tilde{\bar{k}}_i(t)$ depends on the angular coordinates of nodes via $x_{ij}(t)$ in $\tilde{p}(x_{ij}(t))$, but its ``mean-field'' approximation~$\bar{k}_i(t)$ in Equation~(6) does not because it is computed assuming that the angular coordinates are random variables uniformly distributed on~$[0,2\pi]$, and integrating them out. Let $i^{*}$ denote the MLE of the appearance time of node $i$. Using the mean-field approximation $\tilde{\bar{k}}_i(t)\approx \bar{k}_i(t)$ and Equations~(\ref{eq:ml_k_i_t},\ref{eq:k_i_t}), we have that:
\begin{equation}
\label{eq:ml_i}
i^{*} \propto k_i^{-\frac{1}{\beta}}= k_i^{-(\gamma-1)}.
\end{equation}
If $\gamma>1$, Equation~(\ref{eq:ml_i}) implies that the higher the degree of the node, the earlier its MLE appearance time, justifying the following procedure for finding the MLE of the node appearance times in a network with $t$ nodes: sort all nodes in the decreasing order of their degrees $k_1>k_2>\ldots>k_t$, with ties broken arbitrarily, and set their MLE appearance times $i^{*}=1,2,\ldots,t$ in the same order. That is, the node with the largest degree $k_1$ is expected to appear first, $i^*=1$, the second largest degree node $k_2$ appeared second, $i^*=2$, and so on.

From the MLE appearance times of nodes we can compute the MLE of their initial radial coordinates $\{r_i^{*}\}$ as $r_i^{*}=\frac{2}{\zeta}\ln{i^{*}}$, and therefore $\{r_i(j)^{*}\}$, $\forall~j\leq t$ as $r_i(j)^{*}=\beta r_i^{*}+(1-\beta) r_j$, $r_j=\frac{2}{\zeta}\ln{j}$. We now have all the ingredients in place to replay the growth of the network according to E-PSO to find the MLE of the node angular coordinates $\{\theta_i^{*}\}$. We describe this next.

\subsection{HyperMap}
\label{sec:hypermap}

The simple algorithm in Figure~\ref{fig:the_method} fully specifies the HyperMap method. On its input it takes the network adjacency matrix  $\alpha_{ij}$ and the network parameters
$m, L, \gamma, T, \zeta$, and computes radial and angular coordinates $r_i(t), \theta_i$, for all nodes $i \leq t$ in the network.~\footnote{The code implementing HyperMap can be found online at~\cite{hypermap}.} To simplify the notation and the description below we henceforth drop the MLE superscript $^*$ from all variable names.

\begin{figure}
\begin{center}
\begin{minipage}{3.5in}
\begin{algorithm}[H]
\begin{algorithmic}[1]
\small
\STATE Sort node degrees in decreasing order $k_1> k_2>\ldots>k_t$ with ties broken arbitrarily.
\STATE Call node $i$, $i=1,2,\ldots,t$, the node with degree $k_i$.
\STATE Node $i=1$ is born, assign to it initial radial coordinate $r_1=0$ and random angular coordinate $\theta_1 \in [0, 2\pi]$.
\FOR{$i=2$ to $t$}
    \STATE Node $i$ is born, assign to it initial radial coordinate $r_i=\frac{2}{\zeta}\ln{i}$.
    \STATE Increase the radial coordinate of every existing node $j < i$ according to $r_j(i)=\beta r_j +(1-\beta)r_i$.
    \STATE Assign to node $i$ angular coordinate $\theta_i$ maximizing $\mathcal L_2^{i}$ given by Equation (\ref{eq:local_likelihood_2}).
\ENDFOR
\normalsize
\end{algorithmic}
\end{algorithm}
\end{minipage}
\end{center}
\caption{The HyperMap Embedding Algorithm.}
\label{fig:the_method}
\end{figure}

HyperMap first estimates the MLE appearance (or birth) times of nodes $i=1,2,\ldots,t$, as described earlier. We call the node born at time $i$ node $i$. Having a sequence of MLE node birth times, HyperMap replays the hyperbolic growth of the network in accordance with the E-PSO model as follows. When a node is born at time $1 \leq i \leq t$, it is assigned an initial radial coordinate $r_i=\frac{2}{\zeta}\ln{i}$, and every existing node $j < i$ moves increasing its radial coordinate according to $r_j(i)=\beta r_j +(1-\beta)r_i$. The method assigns to a new node $i > 1$ the angular coordinate $\theta_i$ that maximizes its local likelihood $\mathcal L_2^{i}$. This likelihood is a function of $\theta_i$, since $x_{ij}$ depends on $\theta_i$, $p(x_{ij})$ depends on $x_{ij}$, and $\mathcal L_2^{i}$ depends on $p(x_{ij})$.

The maximization of $\mathcal L_2^{i}$ can be performed numerically, by sampling the likelihood $\mathcal L_2^{i}$ at different values of $\theta$ in $[0, 2\pi]$ separated by intervals $\Delta\theta=\frac{1}{i}$, and then setting $\theta_i$ to the value of $\theta$ that yields the largest value of $\mathcal L_2^{i}$. Since, to compute $\mathcal L_2^{i}$ for a given $\theta$ we need to compute the connection probability between node $i$ and all existing nodes $j <i$, we need a total of $O(i^2)$ steps to perform the maximization. If there are $t$ nodes in total, we need $O(t^3)$ running time to map the full network. We note that due to the mean-field approximation leading to Equation~(\ref{eq:ml_i}), and the above discrete sampling of the likelihood, HyperMap is an approximate MLE algorithm.

\textbf{Specifying input parameters.}  Parameter $\zeta > 0$ can be set to any value, so that we set it to $\zeta=1$. Parameter $m$ can be obtained from historical data of the evolution of the network. If such data is available, then $m$ is the average number of connections that nodes have once they first appear in the data. If no historical data are available, $m$ could be set, as an approximation, to the minimum observed node degree in the network. Given the average node degree $\bar{k}$ in the network, and knowing $m$ and $\bar{k}$, we get $L=\frac{\bar{k}-2m}{2}$. The power law exponent $\gamma$ can be obtained from the degree distribution of the network. We have seen (see Section~\ref{sec:modified_pso_model}) that for the AS Internet $m \approx 1.5$,  $L\approx 1$, and $\gamma=2.1$. Finally, as we show in the next section parameter $T$ can be found experimentally. For the AS Internet, we estimate $T\approx 0.8$. We note that HyperMap is a deterministic algorithm: if one fixes in step $3$ of Figure~\ref{fig:the_method} the angular coordinate $\theta_1$ of node $i=1$ to a specific value, then the method will produce the same output in different runs.

\textbf{Correction steps.} The accuracy of HyperMap can be improved by occasionally running a ``correction step''  right after step $7$ in Figure~\ref{fig:the_method}. At each time $i$ that we run a correction step we visit each existing node $j \leq i$, and having fixed the coordinates of the rest of the nodes $l \leq i$, we update its angle to the value $\theta_j'$ that maximizes:
\begin{equation}
\label{eq:local_likelihood_2_correction}
\widetilde{\mathcal L_2^{j}}=\prod_{1 \leq l \leq i} p(x_{jl})^{\alpha_{jl}}\left[1-p(x_{jl})\right]^{1-\alpha_{jl}},~~l\ne j,
\end{equation}
where $x_{jl}$ is the hyperbolic distance between $j$ and $l$ when the youngest of the two nodes appeared, and $p(x_{jl})$ is given by Equation~(\ref{eq:p_x_ji}), using in it
$R_j$ if $j > l$ or $R_l$ if $j < l$.  For improved accuracy, each correction step can be repeated a few times. We have observed that these correction steps are beneficial when run at relatively small times $i$, not exceeding a few hundred nodes. Running them at larger times may not be beneficial, as the accuracy improvement may not be significant enough to justify the longer running times.

\section{Validating the HyperMap}
\label{sec:validation}

\subsection{Basic validation metrics}

To evaluate how well HyperMap maps a given network we use two measures: (i)~how close the empirical connection probability, which is the probability that there is a link between a pair of mapped nodes located at hyperbolic distance $x(t)$, is to the theoretical prediction, i.e., the global connection probability $\tilde{p}(x(t))$ in Equation~(\ref{eq:global_prob}); and (ii)~the Logarithmic Loss, $LL$, a standard metric to evaluate maximum-likelihood inference methods~\cite{Cesa-Bianchi2006}. We discuss these two measures next.

After mapping a network with $t$ nodes we have the radial and angular coordinates $r_i(t), \theta_i$, for all nodes $i \leq t$.  We can compute the hyperbolic distance between every pair of nodes ($\frac{t(t-1)}{2}$ pairs total). Some pairs are connected, some are not. We then bin the range of hyperbolic distances from zero to the maximum distance into small bins. For each bin we find all the node pairs located at the hyperbolic distances falling within the bin. The percentage of connected pairs in this set of pairs is the value of the empirical connection probability at the bin. The closer this empirical connection probability to the theoretical, the more successful the HyperMap is in mapping the network.

The logarithmic loss is defined as $LL=-{\ln{\mathcal L}}$, where $\mathcal L$ is the likelihood. Since maximum-likelihood inference methods operate by maximizing the likelihood, the logarithmic loss is a natural metric of the quality of the results that these methods produce. If the results are good, then the logarithmic loss is small. To quantify how small  is ``small,'' one usually compares $LL$ against the one obtained with random parameter assignments.
In our case, we use $LL$ to quantify the quality of the inference of the node angular coordinates, where $\mathcal L$ is the likelihood $\mathcal L_2$ given by Equation~(\ref{eq:likelihood_2}). That is, we first compute $LL$ using the inferred node coordinates $\{r_i(t), \theta_i\}$, and then compare the result to the case where $LL$ is computed using the inferred $r_i(t)$'s and \emph{random} $\theta_i$'s drawn uniformly from $[0, 2\pi]$. We denote the former by $LL^{inf}$ and the latter by $LL^{rand}$. The smaller the $LL^{inf}$ compared to $LL^{rand}$, the better the quality of the mapping, i.e., the better E-PSO describes a given network.  In particular, the ratio $r_{LL}=e^{-LL^{inf}}/e^{-LL^{rand}}=e^{(LL^{rand}-LL^{inf})}$ is the ratio of the likelihood with the inferred angular coordinates to the likelihood with random angular coordinates. The higher this ratio, the better the mapping quality.

\subsection{Synthetic Networks}
\label{sec:synthetic_nets}

\begin{figure*}
\centerline{
\subfigure[$\gamma=2.1$, $T=0.4$.]{\includegraphics [width=1.8in, height=1.18in]{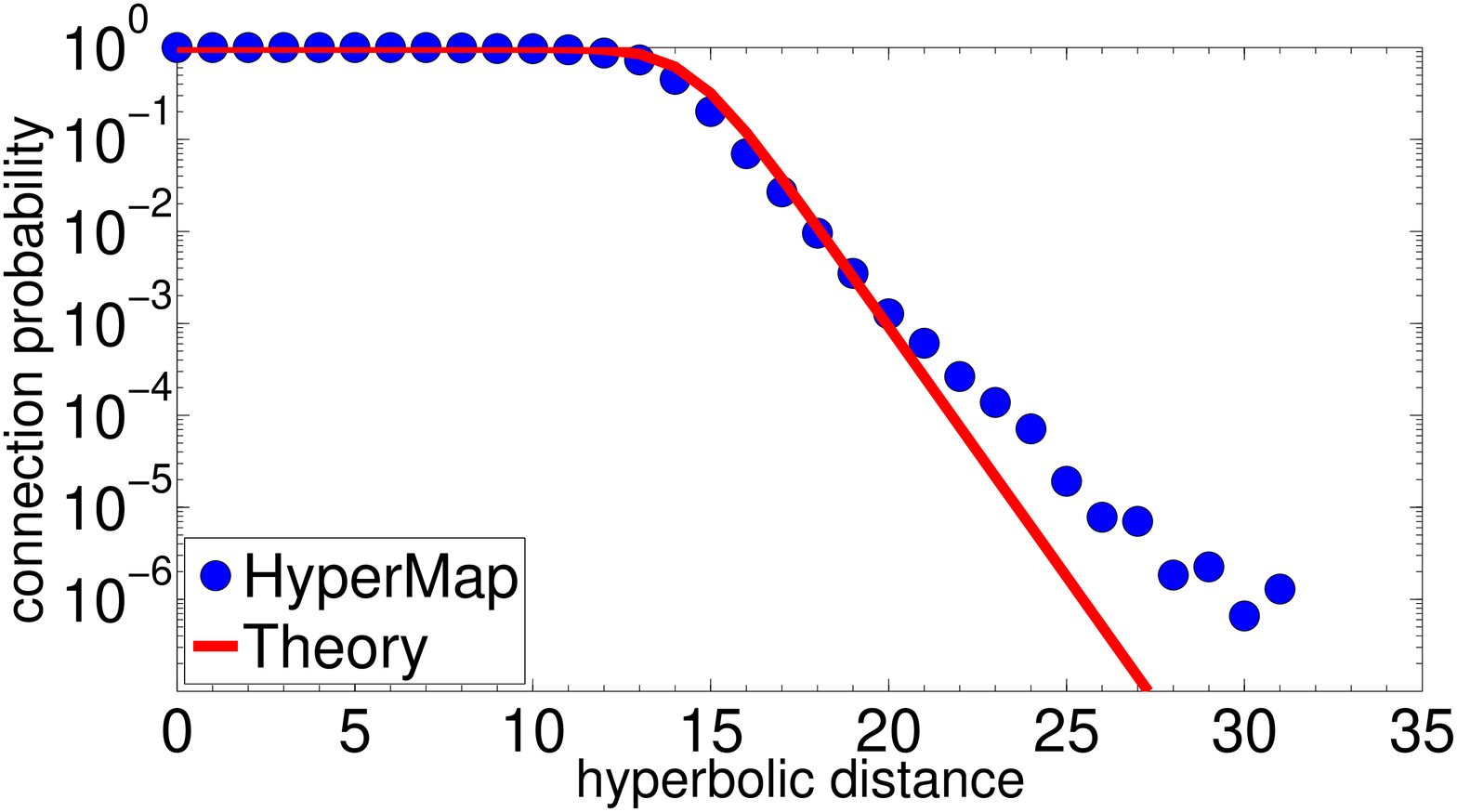}}\hfill
\subfigure[$\gamma=2.1$, $T=0.7$.]{\includegraphics [width=1.8in, height=1.18in]{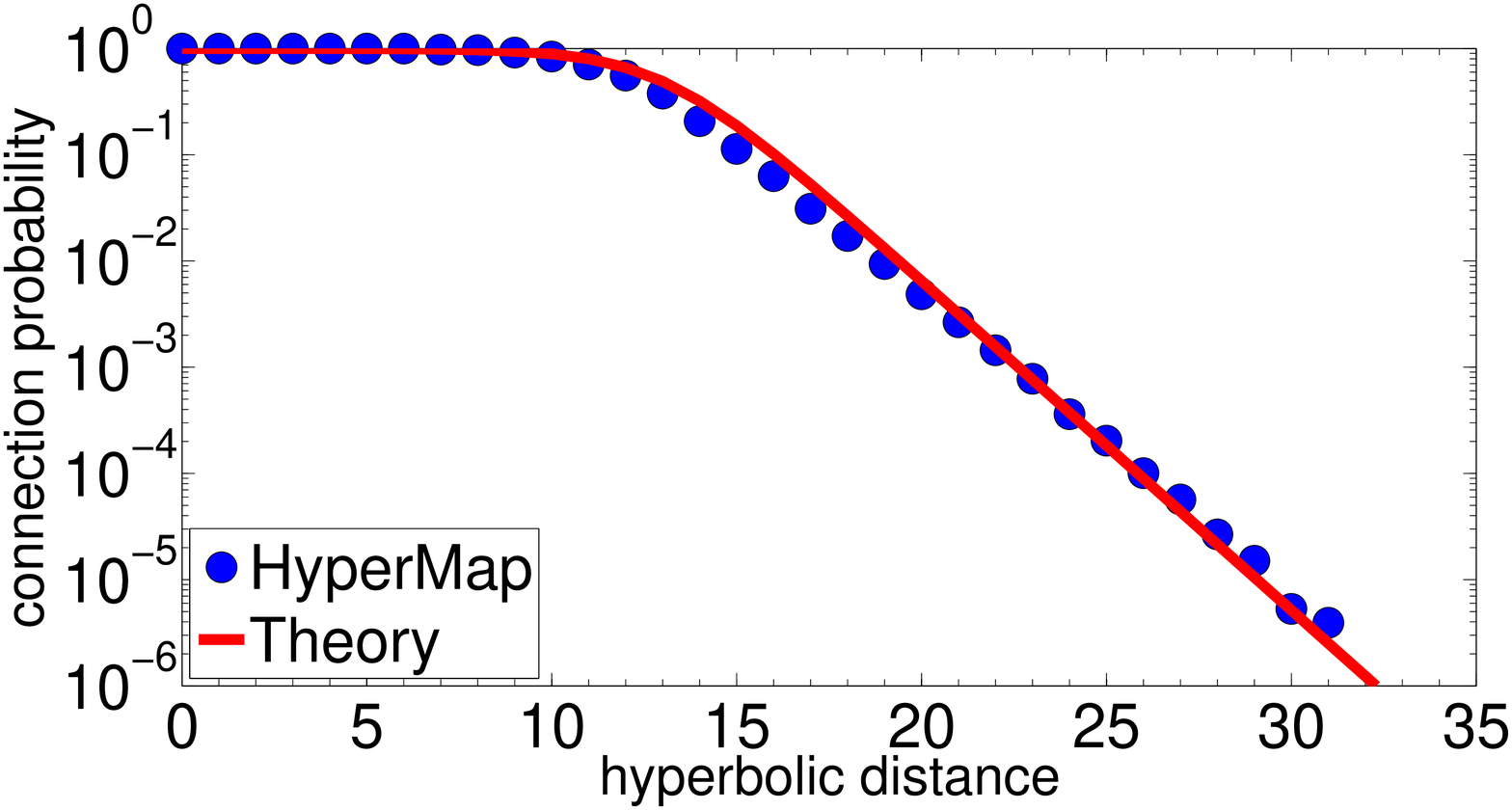}}\hfill
\subfigure[$\gamma=2.5$, $T=0.4$.]{\includegraphics [width=1.8in, height=1.18in]{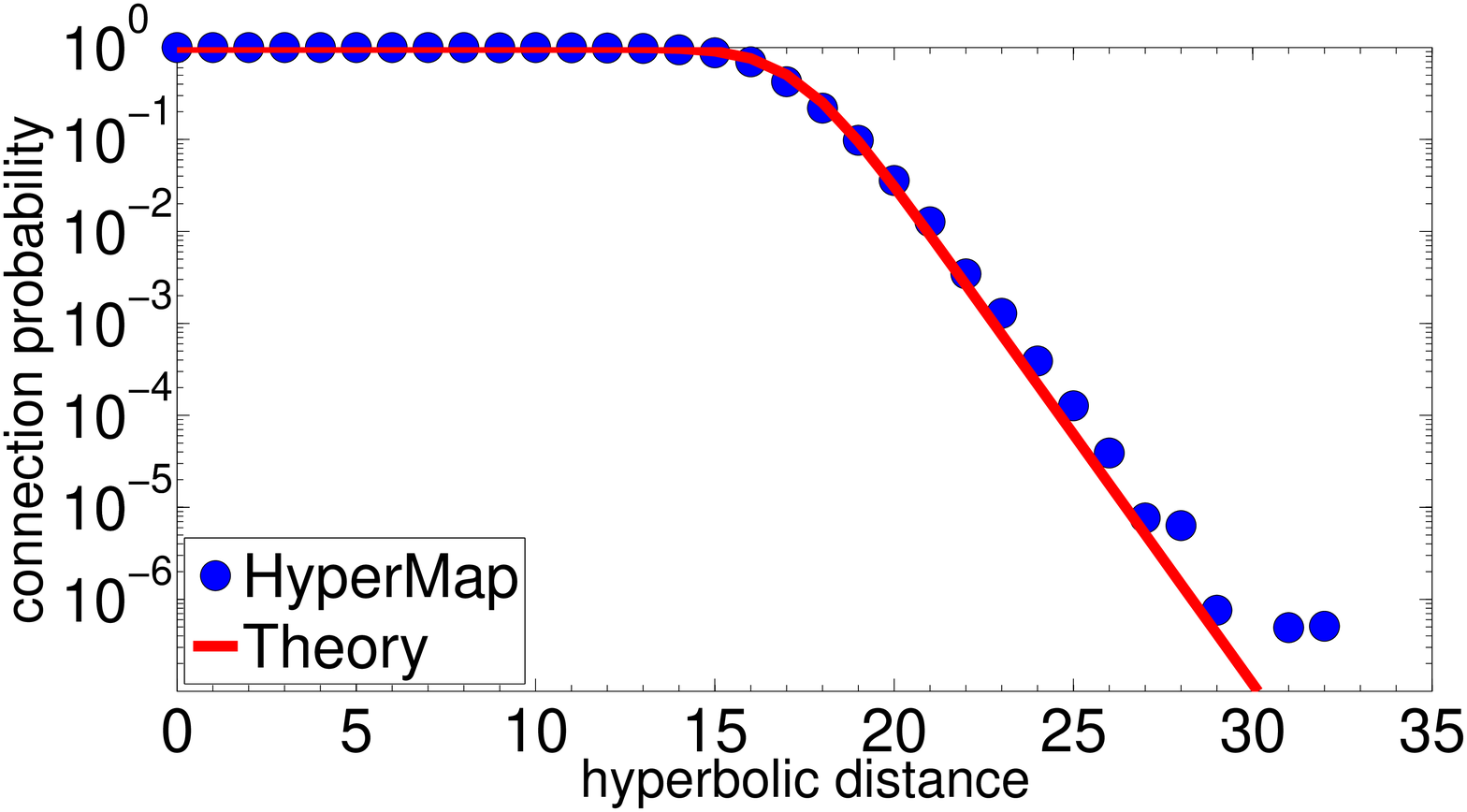}}\hfill
\subfigure[$\gamma=2.5$, $T=0.7$.]{\includegraphics [width=1.8in, height=1.18in]{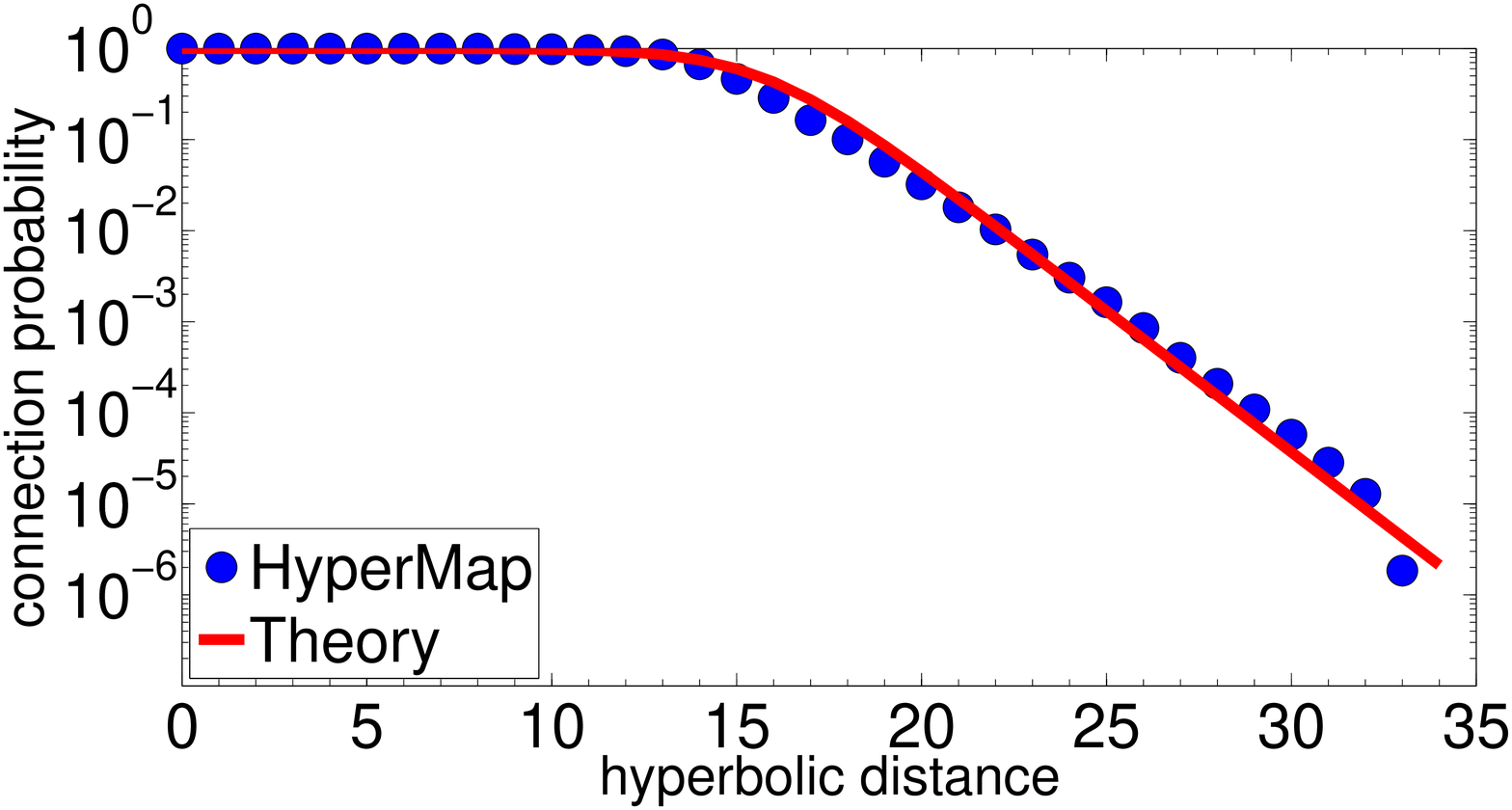}}}
\caption{Connection probability using the inferred node coordinates vs. theoretical prediction given by Equation~(\ref{eq:global_prob}).}
\label{fig:synthetic_1}
\end{figure*}

We first validate HyperMap on synthetic networks, and then apply it to the real AS Internet in the next section. In particular, we first grow synthetic networks according to E-PSO up to $t=5000$ nodes, with $m=1.5$, $L=2.5$, $T=0.4, 0.7$, $\gamma=2.1, 2.5$, and $\zeta=1$. Then, we pass these synthetic networks to HyperMap using their corresponding $m, L, \gamma, T, \zeta$ values, and compute radial and angular coordinates for all the nodes.  HyperMap also runs four correction steps as described in the previous section, right after all nodes with degrees $k \geq 60, 40, 20, 10$ appear in the network. Using the node coordinates given by HyperMap we compute the global connection probability and juxtapose it against the theoretical prediction given by Equation~(\ref{eq:global_prob}).  The results are shown in Figure \ref{fig:synthetic_1}, where for the $x$-axis in the plots (hyperbolic distance) we use bins of size $1$. From the figure, we observe a very good match between the computed connection probability and the theoretical prediction, indicating that HyperMap performs very well.

Table~\ref{tab:synthetic_3} reports the logarithmic losses in the considered networks, as well as the ratio $r_{LL}=e^{(LL^{rand}-LL^{inf})}$. From the table we observe that the logarithmic losses using the inferred angular coordinates ($LL^{inf}$) are significantly smaller than those with random angular coordinates ($LL^{rand}$) and that the ratio $r_{LL}$ is very high.  In the table we also report $LL^{real}$, which is the logarithmic loss if we use the real radial and angular coordinates of nodes. We see that $LL^{inf}$ is very close to $LL^{real}$. We note that HyperMap also performs well if it is applied without correction steps. The corresponding $r_{LL}$ ratios in this case for the networks in Table~\ref{tab:synthetic_3} (from top to bottom) are $r_{LL}=e^{115000}, e^{58000}, e^{176000}, e^{78000}$, which are still quite high. These results show that HyperMap is very accurate at inferring the node coordinates in synthetic networks, suggesting that it may be also accurate in application to real networks.

\begin{table}
\begin{center}
\begin{tabular}{|c|c|c|c|c|}
\hline Network & $LL^{real}$ & $LL^{inf}$ & $LL^{rand}$ & $r_{LL}$\\
\hline $\gamma=2.1, T=0.4$ & $2.4 \times10^{4}$ & $2.9\times10^{4}$ & $17\times10^{4}$ & $e^{141000}$\\
\hline $\gamma=2.1, T=0.7$ & $4.1 \times10^{4}$ &$4.1\times10^{4}$ & $11\times10^{4}$ & $e^{69000}$\\
\hline $\gamma=2.5, T=0.4$ & $3.6 \times10^{4}$ & $3.7\times10^{4}$ & $24\times10^{4}$ & $e^{203000}$\\
\hline $\gamma=2.5, T=0.7$ & $5.6 \times10^{4}$ & $5.8\times10^{4}$ & $15\times10^{4}$ & $e^{92000}$\\
\hline
\end{tabular}
\end{center}
\caption{Logarithmic losses in synthetic networks.
\label{tab:synthetic_3}}
\end{table}

\subsection{Insensitivity to Input Temperature}
\label{sec:T_inference}

\begin{figure}[!ht]
\centerline{
\subfigure[$\gamma=2.1$.]{\includegraphics [width=1.8in, height=1.18in]{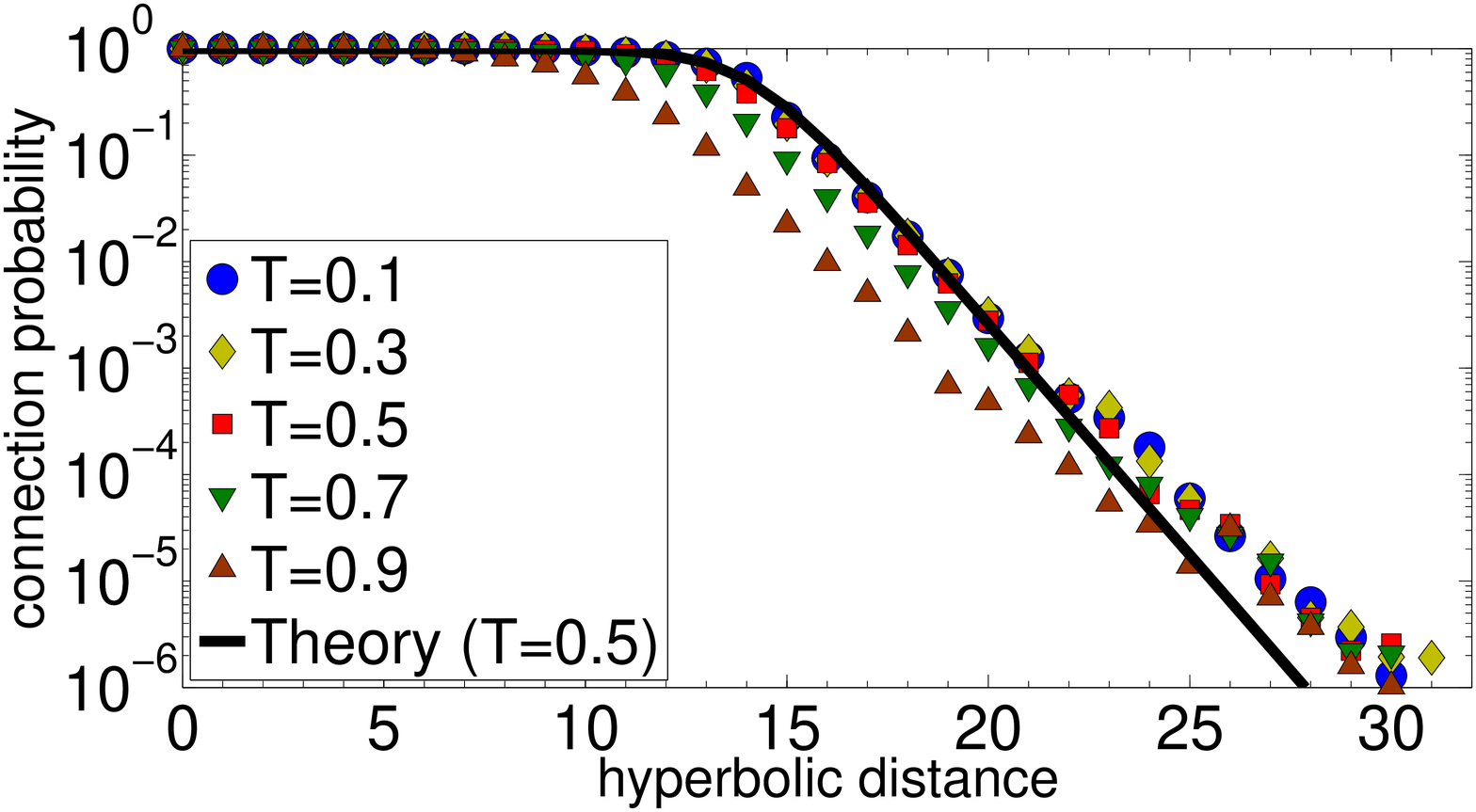}}\hfill
\subfigure[$\gamma=2.5$.]{\includegraphics [width=1.8in, height=1.18in]{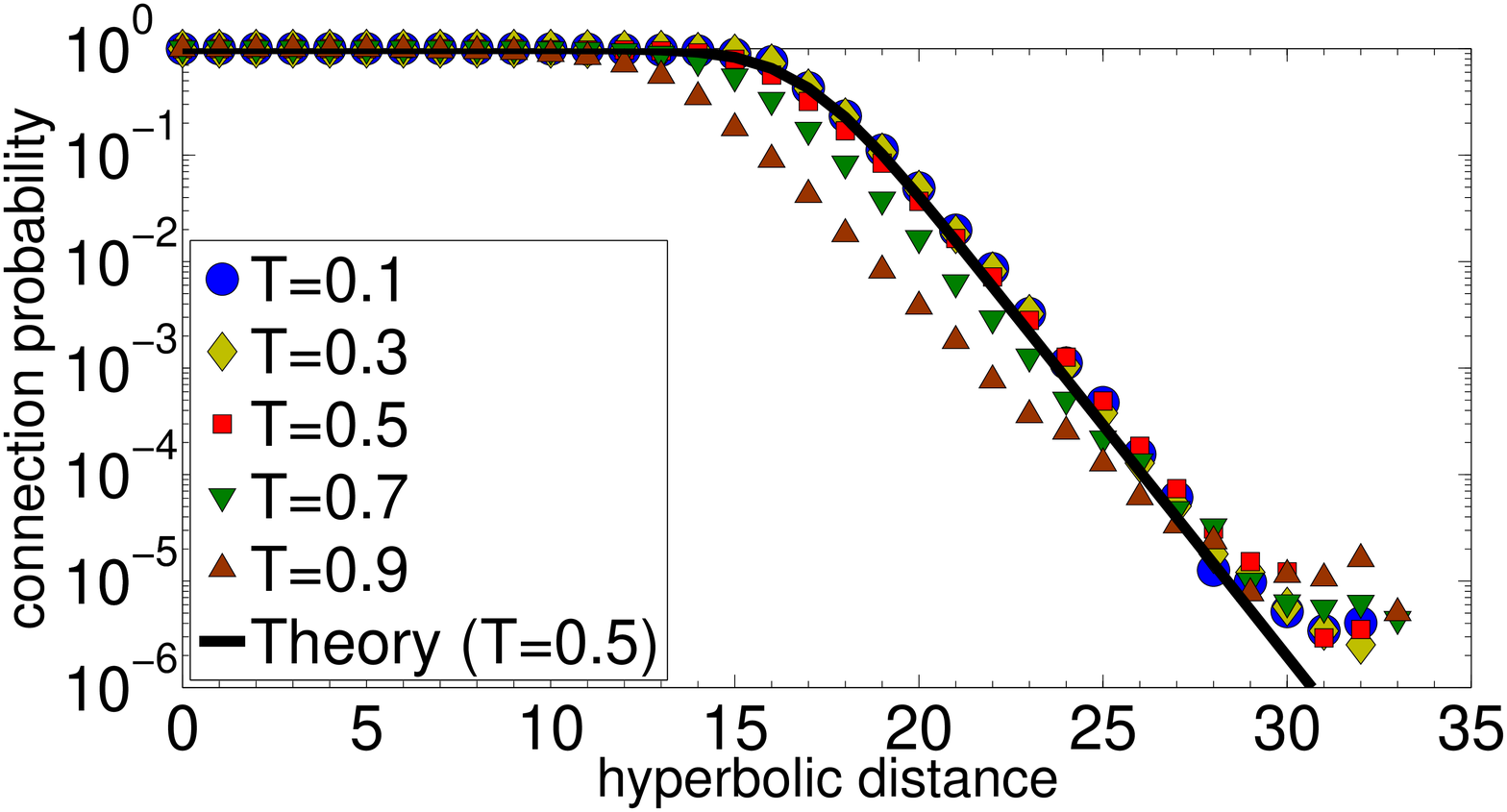}}}
\caption{Insensitivity to input parameter $T$.}
\label{fig:synthetic_4}
\end{figure}

Another important observation contributing to our confidence in HyperMap's accuracy is that it is not too sensitive to the value of the {\em input\/} temperature parameter $T$. To show this we grow synthetic networks using the same parameters as before and $T=0.5 \equiv T^{real}$. We then map these networks using HyperMap with \emph{different input} temperatures $T=0.1, 0.3, 0.5, 0.7, 0.9$, and compute for each case the empirical connection probability. The results are shown in Figure~\ref{fig:synthetic_4}, where we observe that the inferred connection probability is virtually the same for all values of $T \leq T^{real}$, although there are some discrepancies if $T >T^{real}$. This observation implies that HyperMap is good at inferring the {\em real\/} value of temperature in a given network. Whatever value of $T$ we specify on its input, HyperMap infers real $T$, instead of input $T$, which may be wrong or an artifact. Therefore given a network with an unknown temperature parameter $T$, we can infer $T$ by mapping the network using different temperature values until the inferred connection probability converges as in Figure~\ref{fig:synthetic_4}. Then, given a measured value of the tail slope, we can use Equation~(\ref{eq:global_prob}) to find the $T$ value that best matches the theoretical and the inferred connection probabilities. We have followed this approach for the AS Internet yielding $T \approx 0.8$.

\section{Applying HyperMap to the AS Internet}
\label{sec:real_nets}

\begin{figure}[!ht]
\centerline{\includegraphics [width=2.8in, height=1.5in]{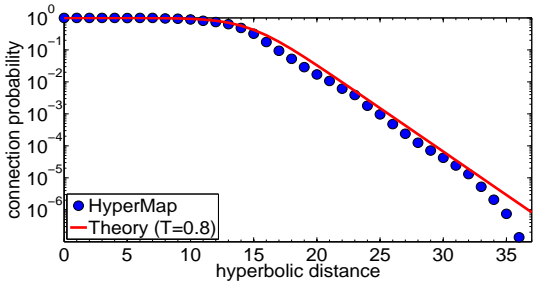}}
\caption{Connection probability in the AS Internet.}
\label{fig:real_1}
\end{figure}

We now consider the AS Internet topology~\cite{ClHy09} described in Section~\ref{sec:modified_pso_model}.
We map the topology using HyperMap as in the previous section using the estimated parameters $m=1.5, L=1, \gamma=2.1, T=0.8$, and $\zeta=1$. As before, we compute the connection probability and Logarithmic Loss (LL). From Figure~\ref{fig:real_1} we observe a remarkable match between the inferred connection probability and the theoretical prediction (Equation~(\ref{eq:global_prob})), while the logarithmic loss is  $LL^{inf}=24\times10^{4}$, and $LL^{rand}=49\times10^{4}$. That is, the $r_{LL}$ ratio is very high, $r_{LL}=e^{(LL^{rand}-LL^{inf})}=e^{250000}$, as in Table~\ref{tab:synthetic_3}. These results indicate that HyperMap performs remarkably well on the AS Internet, too.

\begin{figure*}
\centerline{
\subfigure{\includegraphics [width=2.5in, height=1.5in]{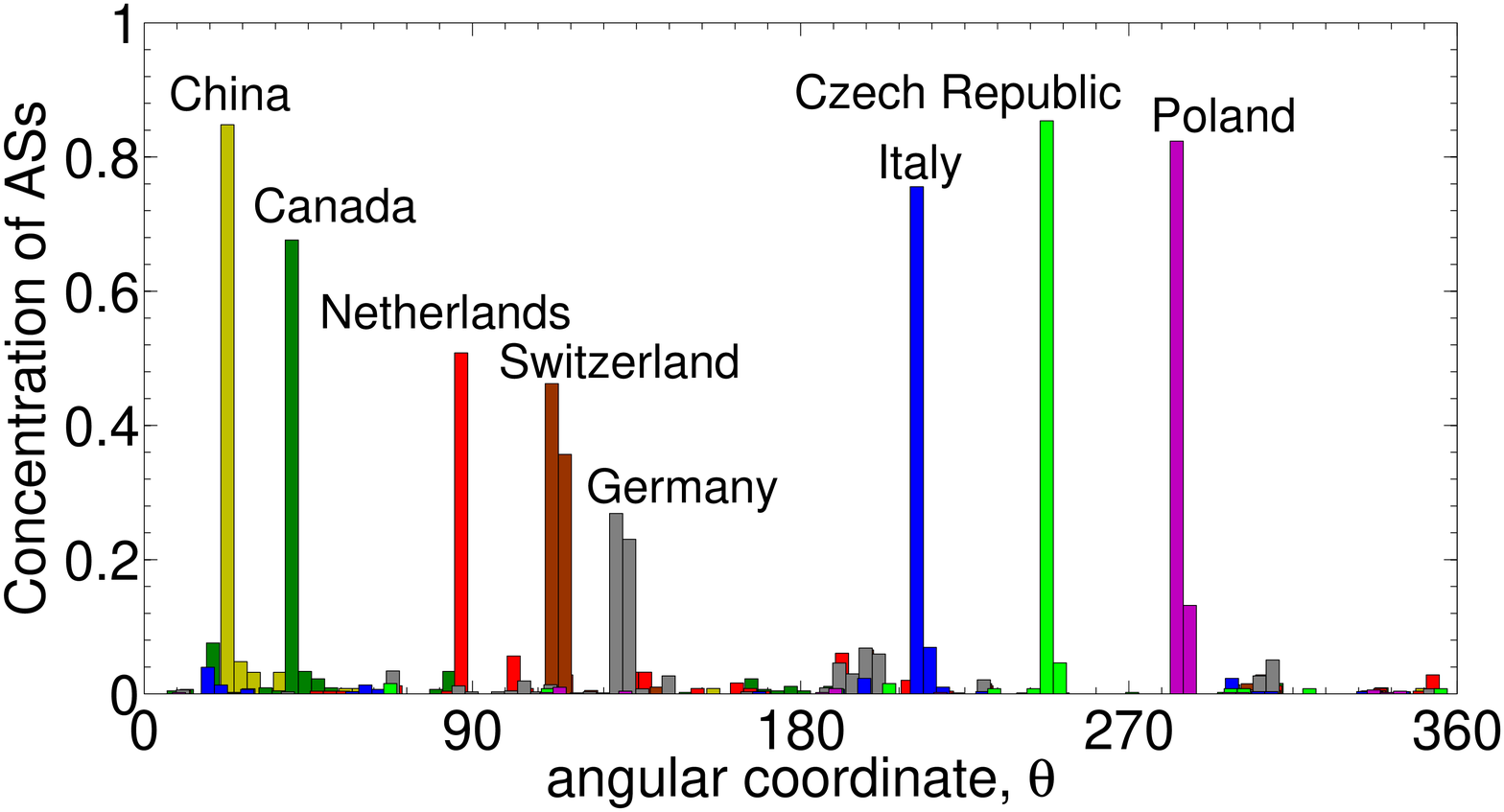}}\hfil
\subfigure{\includegraphics [width=2.5in, height=1.5in]{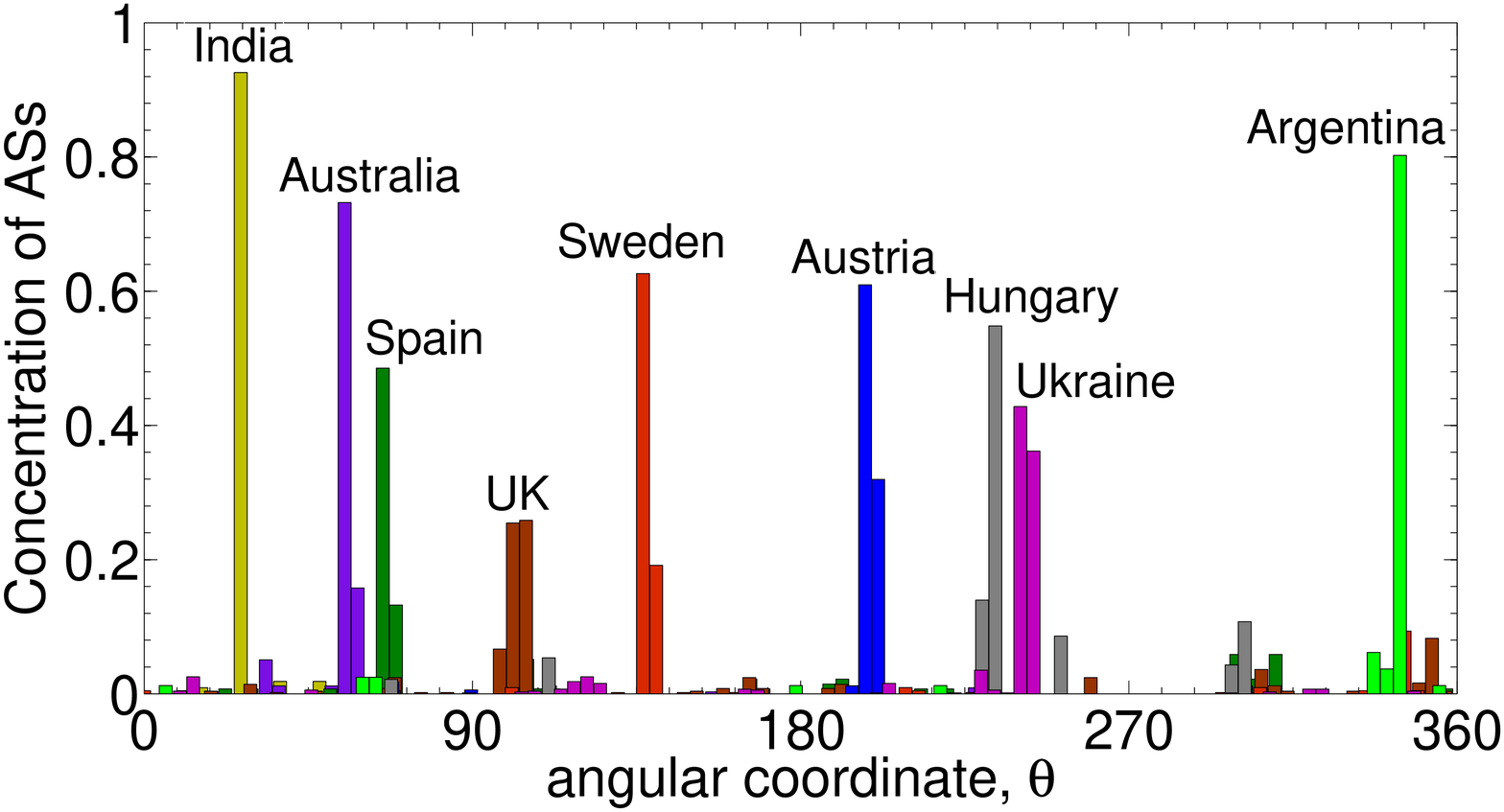}}\hfil
\subfigure{\includegraphics [width=2.5in, height=1.5in]{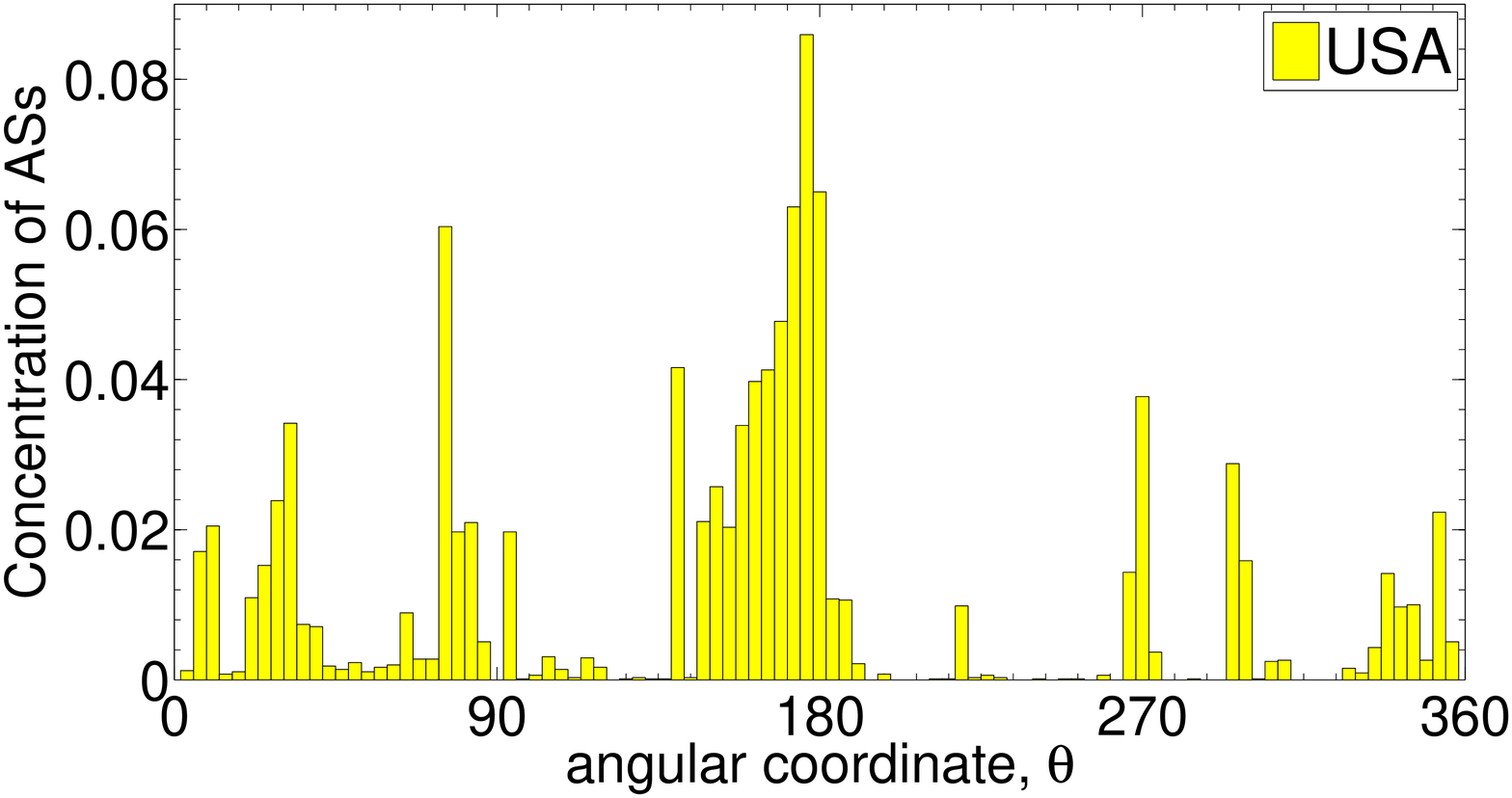}}
}
\caption{Angular distributions of ASs belonging to the same country.}
\label{fig:real_3}
\end{figure*}

In Figure~\ref{fig:real_3}, we also show that the mapping is meaningful, in the sense that HyperMap infers soft communities of ASs belonging to the same country, where by soft communities we mean groups of nodes located close to each other in the space. The figure shows the angular distribution of ASs belonging to the same country for $18$ different countries. The $x$-axis in the plots (angular coordinate) uses bins of size $3.6^{\circ}$. The AS-to-country mapping is taken from the CAIDA AS ranking project~\cite{DiKrFo06}. We observe that even though HyperMap is completely geography-agnostic, it places ASs belonging to the same country close to each other in the angular space. The reason for this is that ASs belonging to the same country tend to connect more densely to each other than to the rest of the world. Connected ASs are attracted to each other, while disconnected ASs repel, and the HyperMap feels these attraction/repulsion forces, placing groups of densely connected ASs in narrow regions, close to each other. As expected, due to significant geographic spread in ASs belonging to the US, these ASs are widespread in $[0^{\circ}, 360^{\circ}]$ as well. We note that other reasons besides geographic proximity may affect the connectivity between ASs, such as economical, political, and performance related reasons. HyperMap does not favor any specific reason but relies only on the connectivity between ASs in order to place the ASs at the right angular (and consequently hyperbolic) distances.

\begin{figure}[!ht]
\centerline{
\subfigure[]{\includegraphics [width=1.85in, height=1.18in]{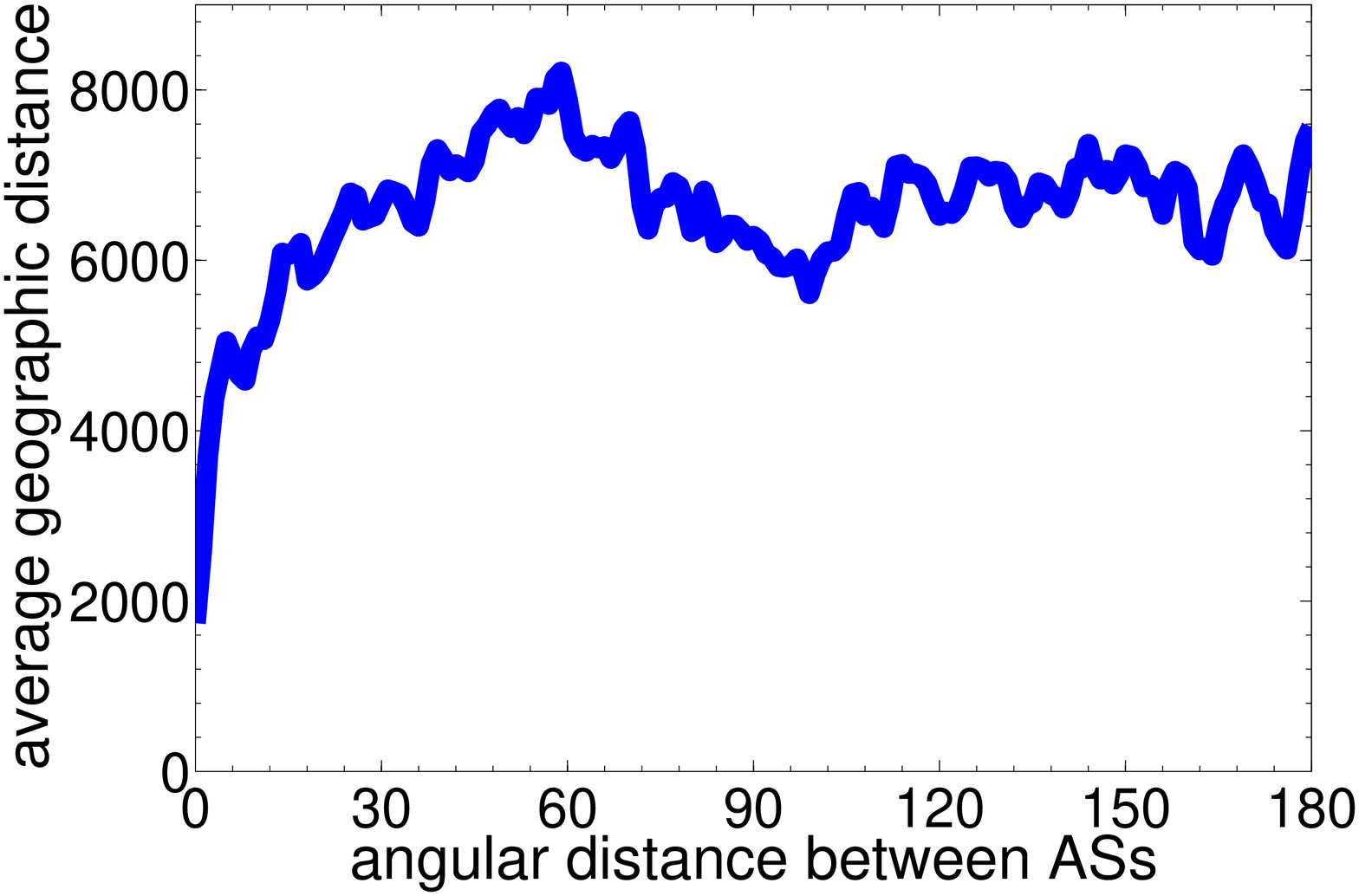}}\hfill
\subfigure[]{\includegraphics [width=1.8in, height=1.18in]{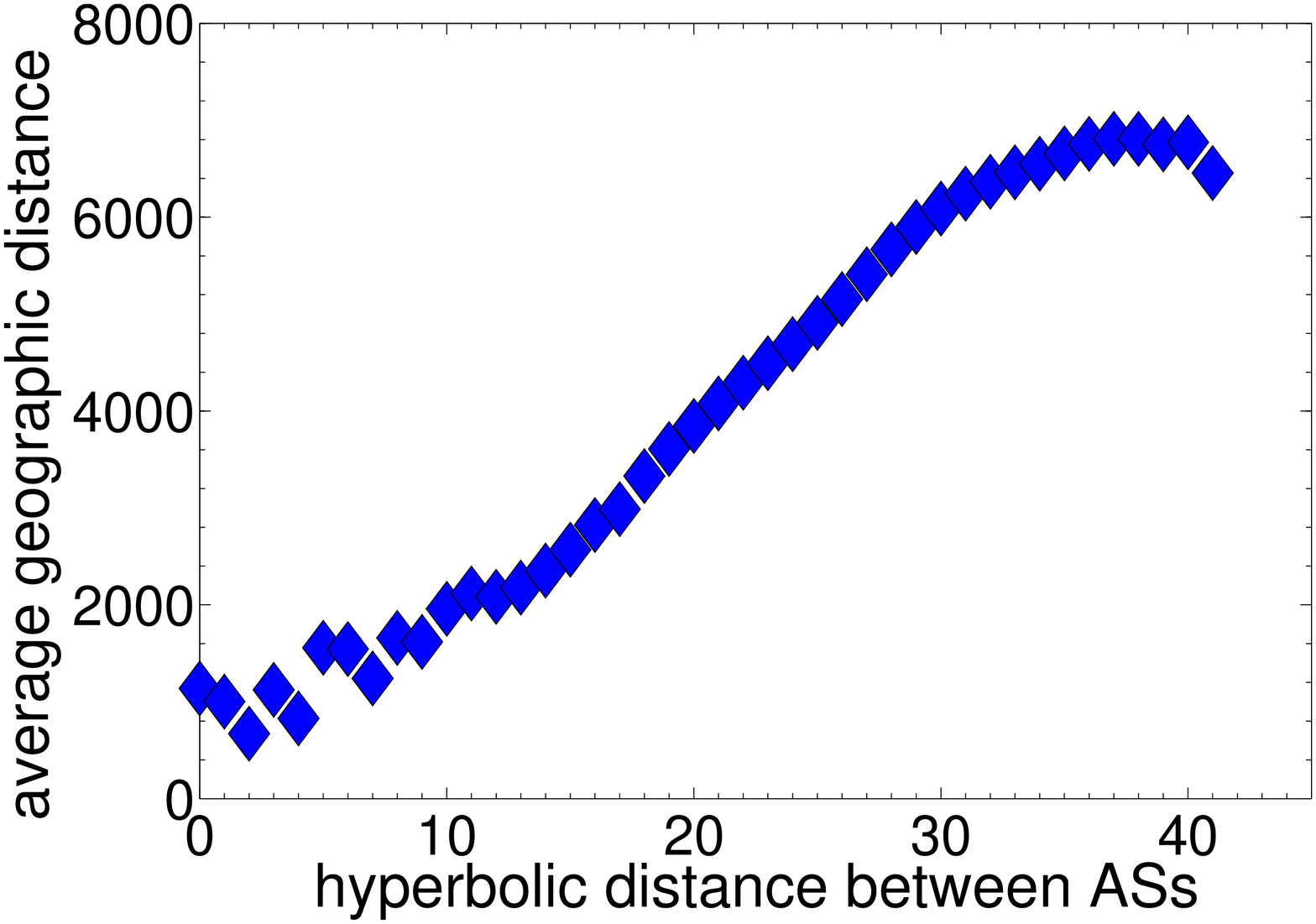}}}
\caption{Average geographic distance (in km) between ASs as a function of their angular distance (plot (a)) and their hyperbolic distance (plot (b)).
\label{fig:H2_vs_Geo}}
\end{figure}

Figure~\ref{fig:H2_vs_Geo}(a) shows the average geographic distance between ASs as a function of their angular distance. We observe that at angular distances below $60^{\circ}$, the average geographic distance tends to grow with the angular distance, which complements Figure~\ref{fig:real_3} confirming that ASs located at smaller angular distances tend to be geographically closer. At large angular distances there is no correlation between geographic and angular distance, because the probability of connections between ASs depends only on their hyperbolic distance, which depends weakly on the angular distance if the latter is large. Figure~\ref{fig:H2_vs_Geo}(b) confirms that the average geographic distance between ASs tends to increase with their hyperbolic distance. Since each AS can span different geographic locations (characterized by their latitudes and longitudes), to draw Figure~\ref{fig:H2_vs_Geo} we first find all the IP prefixes allocated to each AS, geo-resolve them using NetAcuity~\cite{netacuity}, and then compute their center of mass that we use as the AS's geographic coordinates.

Having seen that HyperMap produces an accurate embedding of the AS Internet, in the next section we show that link prediction using this embedding is very efficient, outperforming popular existing link-prediction methods.

\section{Application to Predicting Missing Links}
\label{sec:missing_link_prediction}

\subsection{Background}
\label{sec:background}

Topology measurements of many real networks, not only of the Internet~\cite{LaByCroXie03}, may miss some links. The prediction of missing links is a fundamental problem that attempts to estimate the likelihood of the existence of a missing link between two nodes in a network, based on the observed links and/or the attributes of nodes. See~\cite{LuZhou11} for an in-depth recent survey. Below we recall some basic facts that we need in the rest of the section.

A standard way to evaluate a link prediction technique is to randomly remove a percentage of links from a given network topology, and then work with this incomplete data using the technique to see how well these ``missing,'' i.e., removed links can be predicted~\cite{LuZhou11}.  Formally, consider a network with $t$ nodes and a set $E$ of links between them. Denote by $U$ the set containing all $\frac{t(t-1)}{2}$ possible links. Then, the set of nonexistent links is the set $U - E$. Now, the set $E$ is randomly divided into two parts: the training set, $E^{T}$, which is treated as the known information, and the probe set, $E^{P}$, which is used for testing and no information in this set is allowed to be used for prediction. Clearly $E^{T}\cup E^{P}=E$ and $E^{T}\cap E^{P}=\emptyset$. When a random percentage of links is removed from a network, these missing links are treated as the probe set $E^{P}$, and the remaining links as the training set $E^{T}$.

The standard metric used to quantify the accuracy of a link prediction technique is the \emph{Area Under the Receiver Operating Characteristic Curve (AUC)}~\cite{LuZhou11}.   A link prediction algorithm gives to each non-observed link ($i$, $j$) a score $s_{ij}$ to quantify its existence likelihood.  The better the score of a non-observed link the more likely the link to exist. The prediction algorithm then orders all the non-observed links according to their scores, from the best score to the worst score, with ties broken arbitrarily.  The AUC is the probability that a randomly chosen missing link (i.e., a link in $E^{P}$) is given a better score (i.e., a higher existence likelihood) than a randomly chosen nonexistent link (i.e., a link in $U - E$). The degree to which the AUC exceeds $0.5$ indicates how much better the algorithm performs than pure chance. $\textnormal{AUC}=1$ means a perfect classification (ordering) of the non-observed links, where the missing links are placed in the top of the ordered list.

To get a more detailed characterization of the ability of a technique to predict missing links, the \emph{Receiver Operating Characteristic (ROC)} Curve may also be computed. To compute the ROC Curve we take the ordered set of the non-observed links along with their scores, and consider
each score to be a threshold. Then, for each threshold we calculate the fraction of the missing links that are above the threshold (i.e., the True Positive Rate TPR) and the fraction of the nonexistent links that are above the threshold (i.e., the False Positive Rate FPR). Each point on the ROC curve gives the TPR and FPR for the corresponding threshold. When representing the TPR in front of the FPR, a totally random guess would result in a straight line along the diagonal $y=x$.
The degree by which the ROC curve lies above the diagonal indicates how much better the algorithm performs than pure chance. As the name suggests, the AUC is equal to the total area under the ROC curve.

\subsection{Performance of HyperMap}
\label{sec:mapping_method_performance}

We now check the performance of HyperMap in predicting missing links in the AS Internet topology from Section \ref{sec:real_nets}. We consider the topology consisting of all ASs with degree greater than $2$. We do this to reduce the size of the network we work with to $8220$ nodes. This enables us to compare HyperMap with existing link-prediction techniques, particularly the HRG model and the Katz Index, which are memory-intensive; these techniques require more than $80$GB RAM when applied to the full AS Internet, which is beyond the RAM we have available. We note that HyperMap is not memory-intensive and that the coordinates of nodes with degree $k > k'$ do not depend on the coordinates of nodes with degree $k \leq k'$.

To check HyperMap's performance we \emph{first} remove a percentage $p=10\%, 20\%, 30\%$ of links from the topology, and \emph{then} embed the resulting topology using HyperMap, as described in Section \ref{sec:real_nets}. After the embedding, the score $s_{ij}$ between a disconnected pair of nodes $i, j$, i.e., the score of each non-observed link ($i$, $j$), is the hyperbolic distance $x_{ij}$ between the nodes $i$ and $j$. The smaller this score, i.e., the smaller the hyperbolic distance between the two nodes, the more likely it is that a link between these two nodes is missing, since the connection probability (see Equation~(\ref{eq:global_prob})) is a decreasing function of $x_{ij}$.

\begin{figure}[!ht]
\centerline{\subfigure{\includegraphics [width=2.8in, height=1.5in]{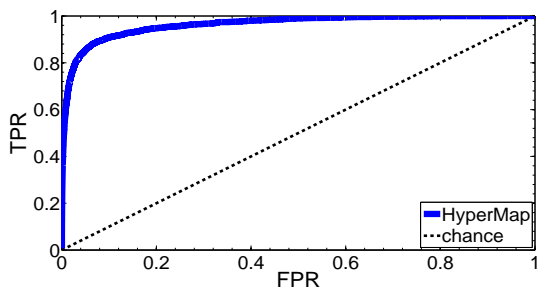}}}
\caption{Receiver Operating Characteristic (ROC) of HyperMap.}
\label{fig:ROC}
\end{figure}

The AUC of HyperMap for $p=10\%, 20\%, 30\%$ missing links is respectively $0.963, 0.962, 0.955$.  That is, the AUC is quite high for all the considered percentages of missing links, indicating that the method has a strong predictive power. For comparison, if we use geographic (instead of hyperbolic) distances between ASs, the corresponding AUC values are significantly lower, $0.758, 0.751, 0.741$.
In Figure~\ref{fig:ROC} we also report the ROC curve of HyperMap when $10\%$ of links are missing. From the figure we see that the curve lies far above the diagonal, which indicates a remarkable power in the method for discriminating missing links from nonexistent links. Similar results hold for the other percentages of missing links, not shown to avoid clutter.

\subsection{Comparison to Classical Link-Prediction Techniques}
\label{sec:performance_comparison}

To provide a deeper insight on the HyperMap performance in predicting missing links, we also consider a
set of classical link prediction methods that have been found to perform well in practice in different studies~\cite{ClMo08, LuZhou11}, and
compare their performance to HyperMap's. In particular, we consider the following five techniques:
(i) Common-Neighbors (CN); (ii) Degree-Product (DP); (iii) Inverse Shortest Path (ISP); (iv) Katz Index (Katz); and (v)
another model-based approach, called the Hierarchical Random Graph (HRG) model~\cite{ClMo08}.

For each technique, we consider the topology of the AS Internet from Section~\ref{sec:mapping_method_performance} with $10\%$ missing links and compute its AUC. Each technique assigns a score $s_{ij}$ to every non-observed link ($i$, $j$) as follows:
\begin{itemize}
\item CN:~~$s_{ij}=|\Gamma(i)\cap\Gamma(j)|$
\item DP:~~$s_{ij}=k_i \times k_j$
\item ISP:~~$s_{ij}=1/l_{ij}$
\item Katz:~~$s_{ij}=\sum_{l=2}^{\infty} \epsilon^{l} \times |\textnormal{paths}_{ij}^{<l>}|$
\item HRG:~~$s_{ij}=p_{ij}$
\end{itemize}
where $\Gamma(i)$ denotes the set of neighbors of node $i$ and $|S|$ is the cardinality of set $S$; $k_i$ denotes the degree of node $i$; $l_{ij}$ is the shortest path between nodes $i, j$; $\textnormal{paths}_{ij}^{<l>}$ is the set of all length-$l$ paths from $i$ to $j$ while $\epsilon$ is a free weight parameter; and $p_{ij}$ is a link existence probability, defined by the hierarchical organization of the network and computed using a Markov Chain Monte Carlo method~\cite{ClMo08}.~\footnote{The code to compute the $p_{ij}$'s according to the HRG model is made publicly available by the authors of~\cite{ClMo08} at \url{tuvalu.santafe.edu/~aaronc/hierarchy/}. We used the code as is without any modifications.}

In all the above methods, the higher the score $s_{ij}$, the more likely a link between nodes $i$ and $j$ exists. In principle one can say that all the methods effectively introduce some measures of node similarity under the assumption that more similar nodes connect more likely. In the first four methods (CN, DP, ISP, Katz), such similarity measures are based on the observable structural characteristics of the network topology. CN assumes that the more common neighbors are between the two nodes, the more likely these nodes are connected; DP models the Preferential Attachment~\cite{Dorogovtsev10-book} mechanism; ISP assumes that the closer the two nodes are in terms of the number of hops between them, the more likely they are connected; while Katz assumes that the greater the number of paths between two nodes the more likely these nodes are connected, and weights the number of paths exponentially based on their length to give shorter paths more weight. For the weight parameter we use an  $\epsilon=0.005$, as in~\cite{Liben-Nowell}. Finally, the last method (HRG) is conceptually closer to our approach, in the sense that the node connection probabilities are not defined by the network topology {\it per se}, but by some ``hidden distances'' (which are hyperbolic distances in our case) that lie ``beneath'' the observable topology.

The results are shown in Table~\ref{tbl:performance_others}. From the table we see that CN yields a high $\textnormal{AUC}=0.95$, which is similar to HyperMap's $\textnormal{AUC}=0.96$. However, CN gives accurate predictions only for node pairs that have common neighbors---its good AUC performance when measured across all node pairs is  not surprising, since $94.6\%$ of the missing links are among nodes with common neighbors.
In contrast, by considering only the node pairs with no common neighbors, which comprise $82\%$ of node pairs, and the missing links only among these pairs, CN yields $\textnormal{AUC}=0.5$, since it assigns the zero score to all such node pairs. That is, CN is as good as pure chance in this case, while HyperMap performs remarkably better yielding $\textnormal{AUC}=0.87$. This result is shown in the hard-links AUC column in Table~\ref{tbl:performance_others}. DP also performs similarly to HyperMap, but DP's performance becomes significantly worse if we consider only node pairs with low degrees. For example, if we consider only pairs of nodes with degrees less than $6$, which comprise 42\% of node pairs, and the missing links only among these pairs, DP gives $\textnormal{AUC}=0.59$, while HyperMap performs significantly better with
$\textnormal{AUC}=0.86$. ISP performs worse than HyperMap, and considering again only node pairs with no common neighbors, we get a lower AUC, $\textnormal{AUC}=0.60$ (vs.\ $\textnormal{AUC}=0.87$ in HyperMap). Katz performs better compared to the rest of the existing techniques we consider. Compared to HyperMap it performs virtually the same when considering all node pairs---particularly, its $\textnormal{AUC}$ is $0.961$ vs. $0.963$ in HyperMap. However, again it performs worse if we consider only node pairs with no common neighbors, having $\textnormal{AUC}=0.77$. Finally, HyperMap performs significantly better than HRG, while the AUC of HRG for node pairs with no common neighbors is only $0.53$.

\begin{table}[!ht]
\begin{center}
\begin{minipage}{\textwidth}\hspace*{1.2cm}
\begin{tabular}{|c|c|c|} \hline
Technique & AUC (all links) & AUC (hard links)\\\hline
CN &0.95 & 0.50\\
HyperMap& 0.96 & 0.87\\\hline
DP &0.94&0.59\\
HyperMap& 0.96 & 0.86\\\hline
ISP &0.88&0.60\\
HyperMap&0.96&0.87\\\hline
Katz & 0.96 & 0.77\\
HyperMap &0.96&0.87\\\hline
HRG &0.65&0.53\\
HyperMap&0.96&0.87\\\hline
\end{tabular}
\end{minipage}
\end{center}
\caption{AUC of Classical Link-Prediction Techniques and Comparison to HyperMap.
\label{tbl:performance_others}}
\end{table}

Summarizing, HyperMap performs remarkably well in predicting missing links in the AS Internet compared to popular existing techniques. Most importantly, while some techniques (CN, DP, Katz) perform similarly in predicting the ``easy-to-predict'' missing links (between high-degree nodes with many common neighbors), they perform worse when it comes to predicting the ``hard-to-predict'' missing links (between low-degree nodes with no common neighbors). In that sense one can say that the measure of similarity (angular distances) between nodes in the PSO model reflects reality more accurately than these other approaches do, and that HyperMap is accurate at inferring these similarity distances in the real Internet.

In fact it has been pointed out that the performance of link-prediction heuristics such as CN or Katz applied to real networks can be explained by the existence of latent spaces underlying these networks~\cite{sarkar11}. These spaces, which we call {\em hidden metric spaces}~\cite{SeKrBo08}, impose certain bounds on the hidden distances and connection probabilities between nodes, in particular between disconnected nodes, explaining why CN or Katz perform well. HyperMap performs better because it not only respects the same bounds since it is explicitly based on a latent-space network model (E-PSO), but it also infers accurately these hidden spatial distances between all nodes in the network.

\section{Application to Network Navigation}
\label{sec:navigation}

Finally we show that the HyperMap-inferred map of the Internet is highly navigable. A network embedded in a geometric space is considered \emph{navigable}~\cite{BoKrKc08} if one can perform efficient \emph{greedy routing (GR)} on the network using the node coordinates in the underlying geometric space. In GR, a node's address is its coordinates in the space, and each node knows only its own address, the addresses of its neighbors, and the destination address written in the packet. In its simplest form, GR forwards a packet at each hop to the neighbor closest to the destination in the geometric space, and drops the packet if the current hop is a local minimum, meaning that it does not have any neighbor closer to the destination than itself. In a slightly modified form, which yields better results, GR excludes the current hop from any distance comparisons, and finds the neighbor closest to the destination. The packet is dropped only if this neighbor is the same as the packet's previous hop.

In our case, the underlying geometric space is hyperbolic, and a node's address is its hyperbolic coordinates ($r,\theta$). Here we evaluate the efficiency of GR in synthetic networks constructed according to the E-PSO model, using both the HyperMap--inferred and the real node coordinates. We also report  its efficiency in the AS Internet using the HyperMap--inferred coordinates. We use the modified version of the GR algorithm.

To evaluate the efficiency of GR, one usually uses two metrics~\cite{BoKrKc08}: (i) the percentage of successful paths, $p_s$, which is the proportion of paths that reach their destinations; and (ii) the average hop-length $\bar{h}$ of the successful paths. Table~\ref{tab:navigation} shows the results for the synthetic networks considered in Section~\ref{sec:synthetic_nets}, and for the AS Internet of Section~\ref{sec:real_nets}. From the table, we make several interesting observations. First, from the numbers in parentheses, which correspond to GR's performance in synthetic networks using the real node coordinates, we observe that the E-PSO networks are remarkably navigable, yielding high $p_s$'s and low $\bar{h}$'s. This efficiency is very similar to the one of the non-growing synthetic networks considered in~\cite{KrPa10}, and it is due to the congruency between scale-free network topology and hyperbolic geometry~\cite{KrPa10}.  Second,  from the table we see that in both the synthetic networks and in the real AS Internet, GR's performance using the HyperMap--inferred node coordinates is remarkably high, yielding in all cases success ratios $p_s > 90\%$, while maintaining low path lengths, i.e., low stretch defined as the average ratio of path length to the shortest-path length. Finally, in the synthetic networks, we observe that GR with the HyperMap--inferred node coordinates yields better $p_s$'s compared to GR with the actual node coordinates (numbers in parentheses), especially for the higher temperatures $T$.
The reason for this is that HyperMap always estimates the node coordinates that best fit a given network. Due to randomness in the network construction process, some nodes might have coordinates that deviate from their best-fit values. Such deviations are minimized at $T \to 0$, in which case the connection probability in Equation~(\ref{eq:p_x_ji}) becomes the step-function $p(x_{ij}) \to 1$ if $x_{ij} \leq R_i$, and $p(x_{ij}) \to 0$ if $x_{ij}  > R_i$. We note that the results in Table~\ref{tab:navigation} correspond to applying HyperMap with correction steps, as described in Section~\ref{sec:synthetic_nets}. HyperMap without correction steps still yields good results. In the synthetic networks, $p_s$ ranges from $0.87$ to $0.90$, and $\bar{h}$ ranges from $3.5$ to $4.85$, while in the Internet, $p_s=0.87$ and $\bar{h}=4.00$.

\begin{table}[!ht]
\begin{center}
\begin{tabular}{|c|c|c|}
\hline Network & $p_s$&  $\bar{h}$\\
\hline $\gamma=2.1, T=0.4$ & $0.97~(0.94)$ & $3.33~(3.24)$\\
\hline $\gamma=2.1, T=0.7$ & $0.93~(0.77)$ &$3.71~(3.51)$ \\
\hline $\gamma=2.5, T=0.4$ & $0.97~(0.94)$ & $3.77~(3.69)$\\
\hline $\gamma=2.5, T=0.7$ & $0.91~(0.79)$ & $4.40~(4.17)$\\
\hline AS Internet &$0.92$&3.97\\
\hline
\end{tabular}
\end{center}
\caption{Success ratio $p_s$ and average hop-length $\bar{h}$ of greedy paths using the HyperMap-inferred node coordinates. Numbers in parentheses show the results using the real coordinates.
\label{tab:navigation}}
\end{table}

\section{Discussion and Conclusion}
\label{sec:discussion}

\begin{figure}[!ht]
\centerline{
\subfigure[]{\includegraphics [width=1.8in, height=1.3in]{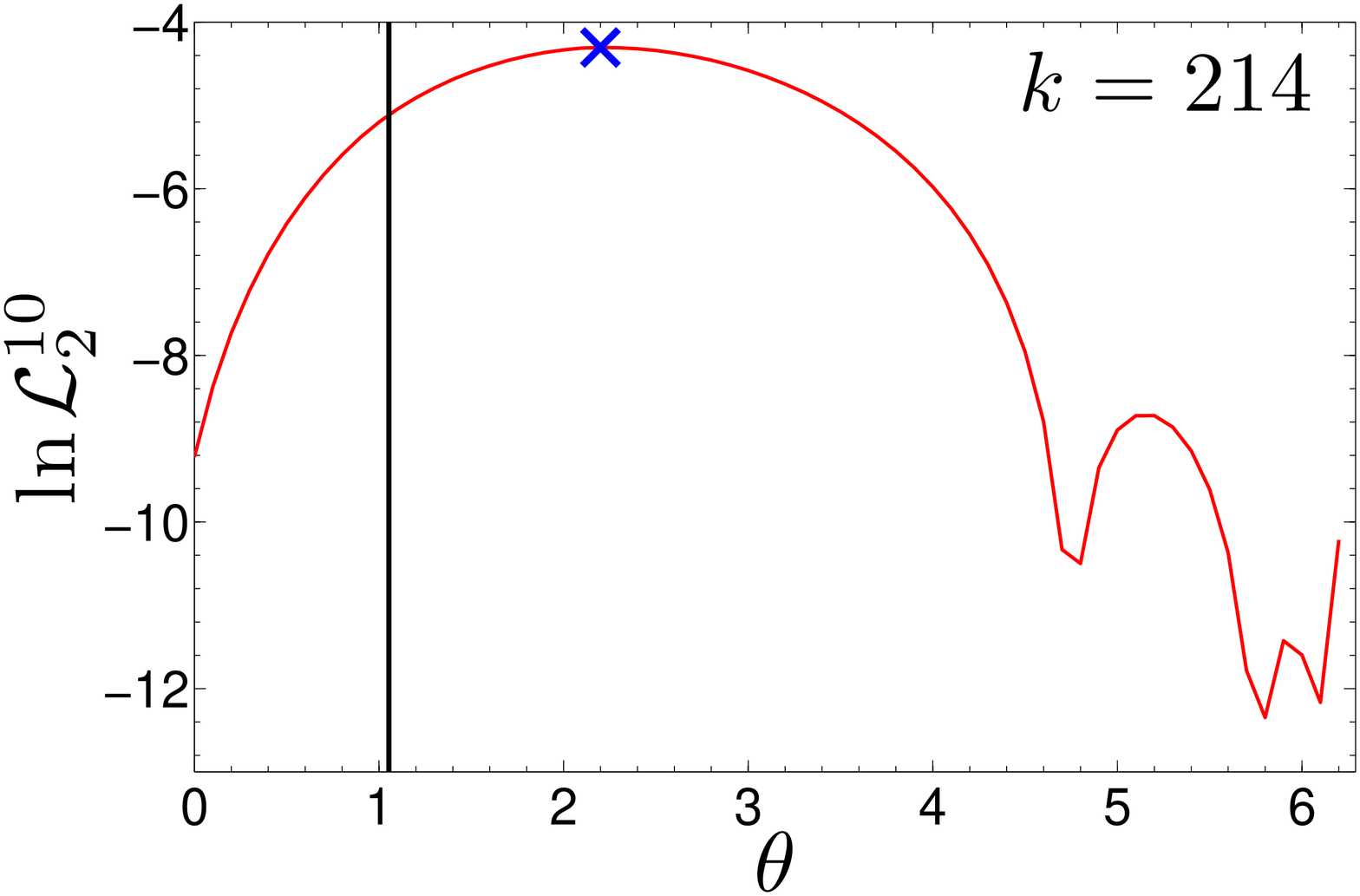}}
\subfigure[]{\includegraphics [width=1.8in, height=1.3in]{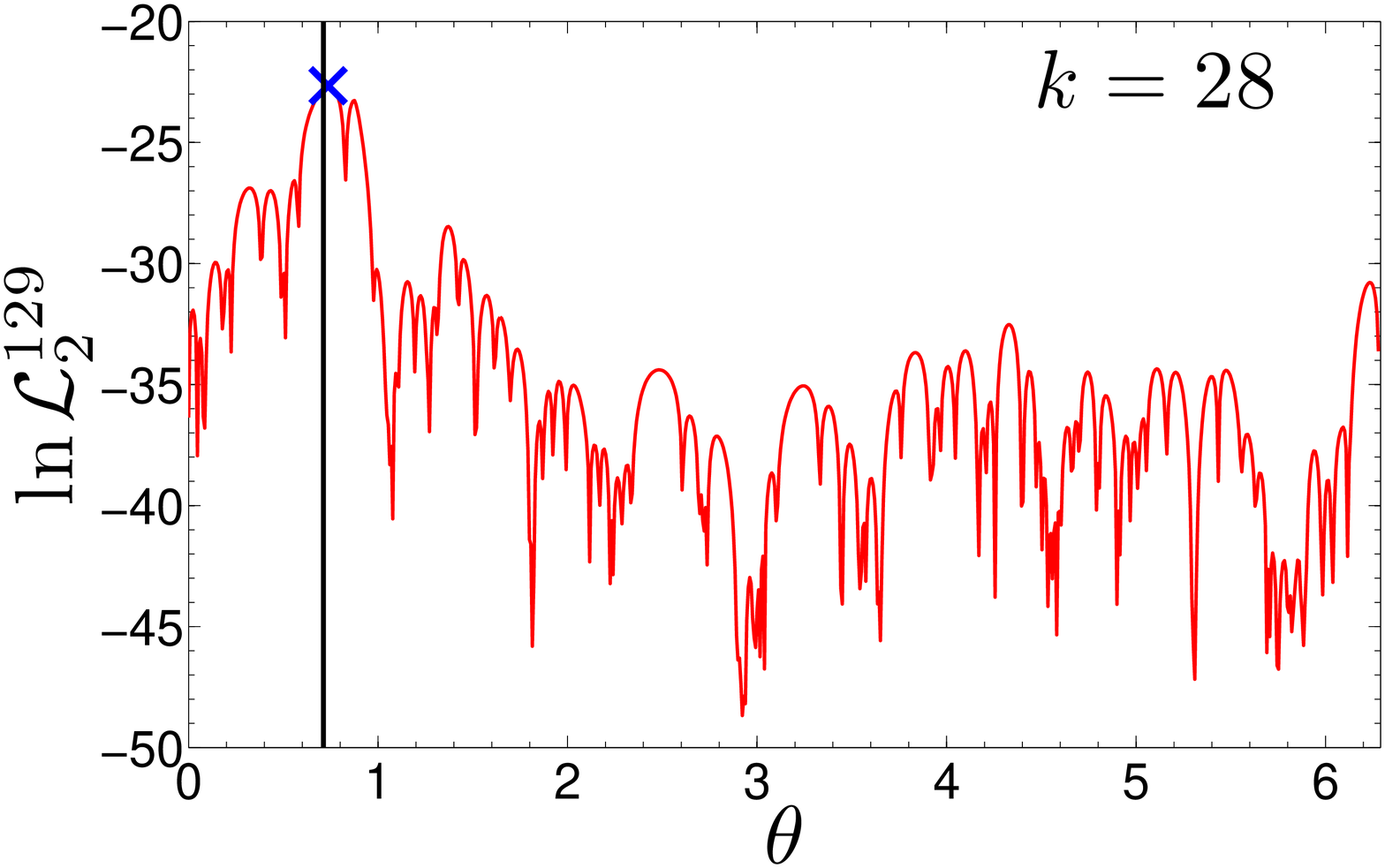}}}
\centerline{
\subfigure[]{\includegraphics [width=1.8in, height=1.32in]{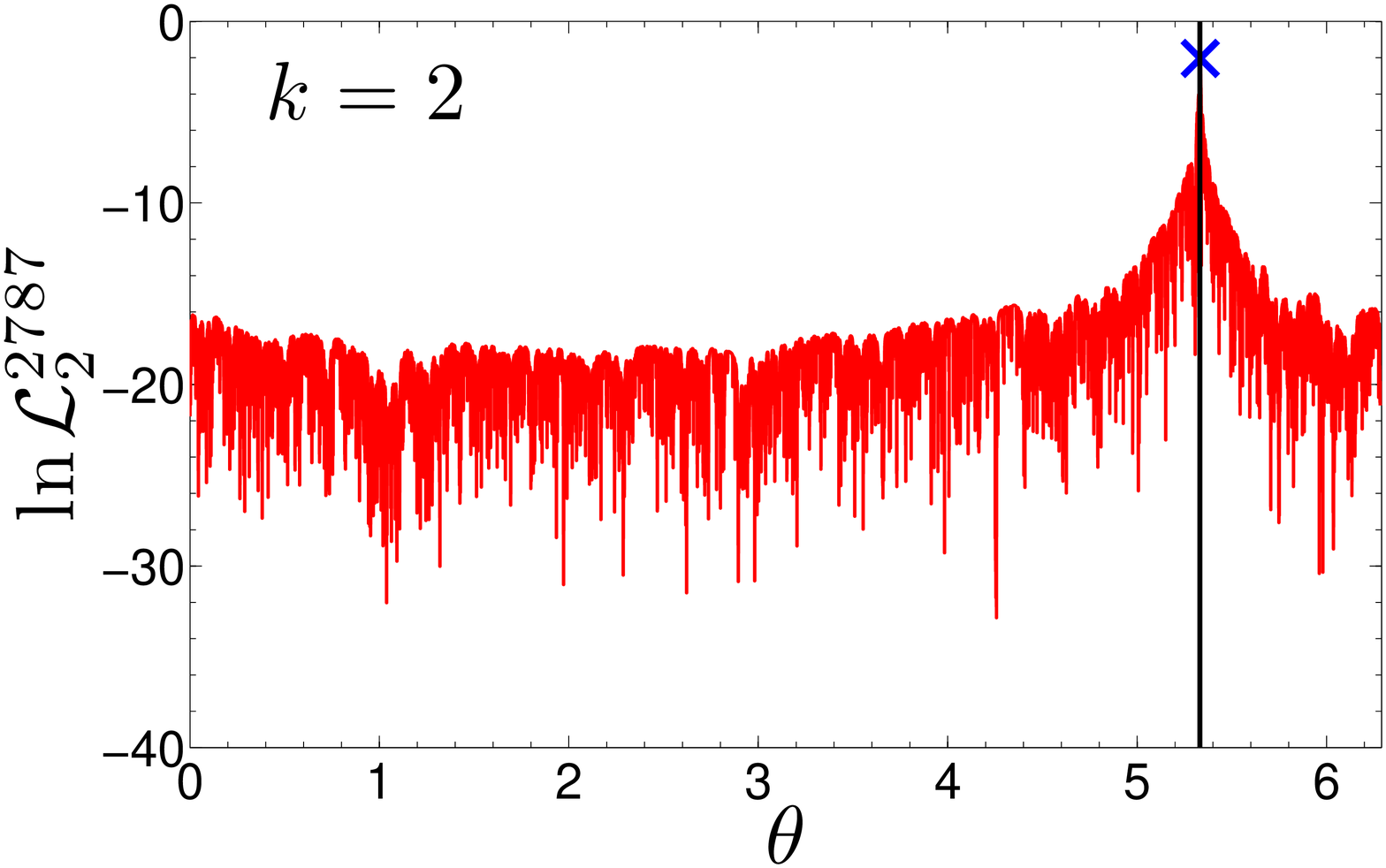}}
\subfigure[]{\includegraphics [width=1.8in, height=1.4in]{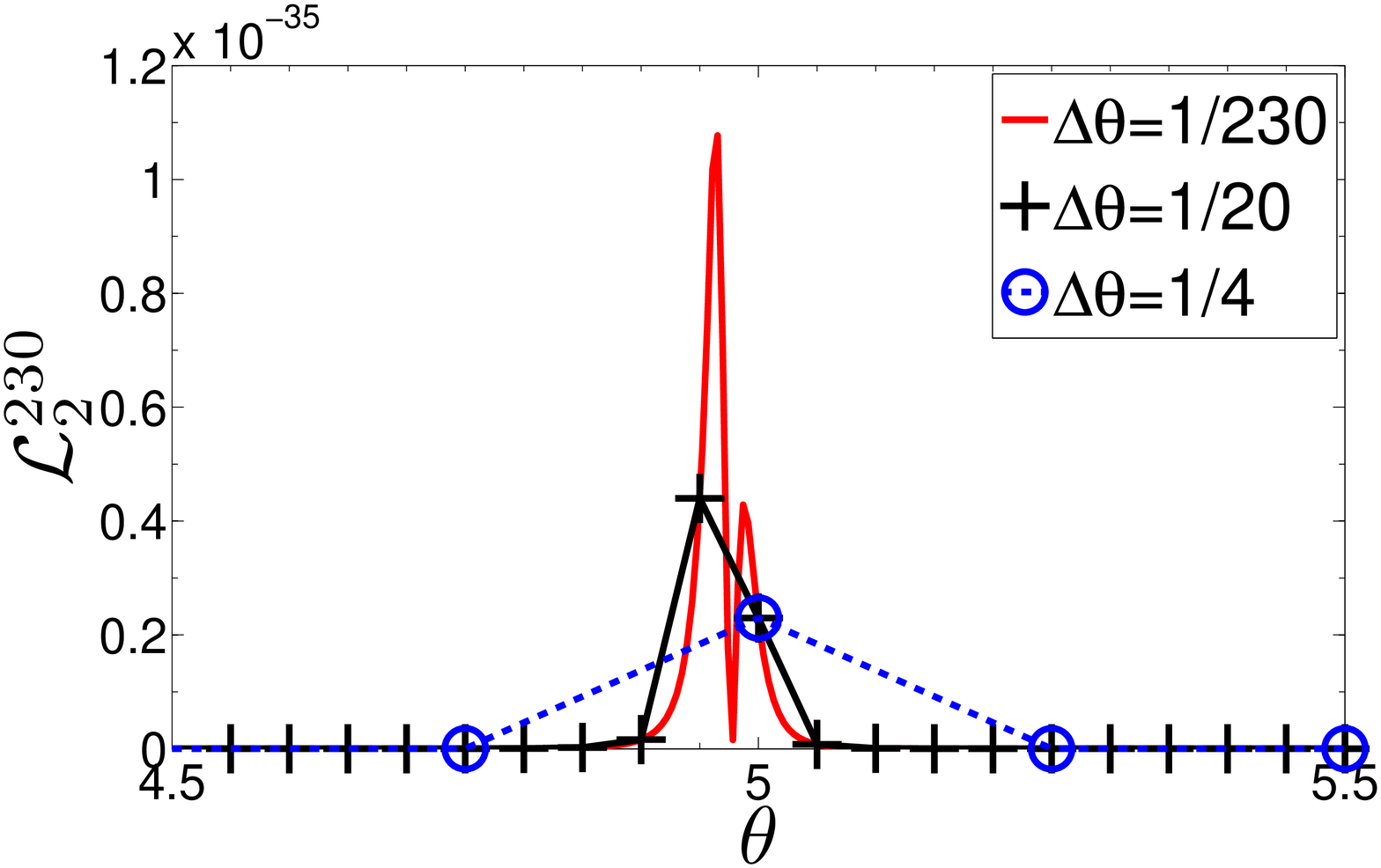}}}
\caption{Local likelihood landscapes for different nodes in a synthetic network and the Internet. Plots~(a), (b), (c) show the log-likelihood $\ln{\mathcal L_2^{i}}$ in Eq.~(\ref{eq:local_likelihood_2}) as a function of the angular coordinate~$\theta$ (in radians) of a given node. The plots correspond to nodes with degrees $k=214, 28, 2$ appearing at MLE times $i=10, 129, 2727$, respectively. The vertical line in each plot shows the real angle of each node in a synthetic network, while the cross shows the angle inferred by HyperMap. By the HyperMap definition, this angle always corresponds to the global maximum of~$\ln{\mathcal L_2^{i}}$. Figure (d) shows the likelihood  $\mathcal L_2^{i}$ for the node appearing at MLE time $i=230$ in the Internet embedding. The likelihood is shown for the range of $\theta \in [4.5, 5.5]$, where it achieves its maximum value. The maximum with the $\theta$-space sampling interval $\Delta\theta=1/230, 1/20, 1/4$ is achieved at $\theta=4.96, 4.95, 5$, respectively.}
\label{fig:theta_sampling}
\end{figure}

Even though we have seen that HyperMap is overall remarkably accurate and efficient, there are aspects of the method that are open for improvement.
One such aspect is the exact estimation of the angular coordinates of the first few nodes appearing at early MLE times.
Specifically, from Equation~(\ref{eq:m_i}), all nodes $i$ for which  $\bar{m}_i(t) \geq i-1$ are all connected to each other with high probability, cf.~Figure~\ref{fig:inet_vs_sim}(f), making it difficult for the method to accurately estimate the exact angular coordinates of such nodes since large zones of different angular coordinates are all quite likely, see Figure~\ref{fig:theta_sampling}(a). However, the number of nodes that have this property is very small, and this effect does not significantly influence the overall efficiency of HyperMap.  For instance, in the synthetic networks considered in Section~\ref{sec:synthetic_nets}, relation $\bar{m}_i(t) \geq i-1$ holds only for the first $33$ nodes when $\gamma=2.1$ and for the first $21$ nodes when $\gamma=2.5$, while for the AS Internet (Section~\ref{sec:real_nets}) it holds only for the first 38 nodes. To illustrate, we consider the $\gamma=2.1, T=0.7$ synthetic network from Section~\ref{sec:synthetic_nets}, and show in Figures~\ref{fig:theta_sampling}(a-c) the log-likelihood $\ln{\mathcal L_2^{i}}$ in Eq.~(\ref{eq:local_likelihood_2}) for nodes appearing at MLE times $i=10, 129, 2727$, having degrees $k=214, 28, 2$. In each case, the angular coordinates of the old nodes $j < i$ are fixed to their real values. From the figures, we observe that when $i=10$ the inference is not exact (Figure~\ref{fig:theta_sampling}(a)), while it becomes extremely precise as $i$ increases (Figures~\ref{fig:theta_sampling}(b),(c)). Similar results hold for the rest of the networks we considered, and for other network parameter values. An interesting open question is whether the method could be improved to infer the angular coordinates of the first few nodes exactly, and whether this improvement would have any significant effects on the overall performance of HyperMap.
The correction steps discussed in Section~\ref{sec:hypermap} are aiming at this direction, by trying to recompute improved angles for the first nodes, considering not only the connections to their previous nodes, but also connections to nodes that appear later. However, they still cannot guarantee that
the inference of these angles will be exact.

Another aspect that is open for improvement is the way the maximization of $\mathcal L_2^{i}$ is performed. As explained in Section~\ref{sec:hypermap}, HyperMap samples the likelihood $\mathcal L_2^{i}$ of every new node $i$ to find the angle $\theta_i$ that maximizes $\mathcal L_2^{i}$. Since the sampling of $\mathcal L_2^{i}$ is done at discrete intervals $\Delta \theta_{i}=\frac{1}{i}$, there might be cases that the true global maximum of $\mathcal L_2^{i}$ is missed. For example, Figure~\ref{fig:theta_sampling}(d) shows $\mathcal L_2^{i}$ for the node appearing at MLE time $i=230$ in the AS Internet embedding, when this $\mathcal L_2^{i}$ is sampled with different $\Delta \theta$ intervals. We see that even though the three sampling intervals yield approximately the same angular coordinate, the likelihood is one order of magnitude larger at  $\theta=4.96$, which is discovered only when $\Delta\theta=\frac{1}{230}$. We thus see that the maximization of $\mathcal L_2^{i}$ is not a trivial issue. In general, decreasing the sampling interval may increase the accuracy of the method but will also increase its running time. We have found that $\Delta \theta_{i}=\frac{1}{i}$ is sufficient  to yield good results in practice, as also illustrated in Figures~\ref{fig:theta_sampling}(b),(c). Further, notice from Figures~\ref{fig:theta_sampling}(a-c) that the likelihood profile becomes abundant with deep local maxima as $i$ increases, justifying the need for the increasingly smaller sampling interval. The correction steps discussed in Section~\ref{sec:hypermap} are also beneficial in this aspect, since they resample the likelihood of a node (Eq.~(\ref{eq:local_likelihood_2_correction})) at future times $i$,  where $\Delta \theta_{i}=\frac{1}{i}$ is smaller. More sophisticated techniques~\cite{optimization_book} that numerically find the global maximum of a function may yield improved performance.  Finding the most efficient option, yielding an adequate balance between computational complexity and embedding accuracy, is another open research problem.

In~\cite{BoPa10}, we have focused on greedy routing, and showed how the AS Internet topology can
be embedded into the hyperbolic plane, by maximizing the likelihood that the topology is produced by the model
of static complex networks from~\cite{KrPa10}. To do so, a localized Metropolis-Hastings algorithm was used, in conjunction with some sophisticated heuristics to guide the algorithm to produce good results in a reasonable amount of compute time. The procedure required manual intervention, such as manually determining good degree thresholds that define layers of nested subgraphs~\cite{BoPa10}. In this paper, we have followed a different approach. We have shown how to embed the AS Internet (and in general, a scale-free network) by replaying its  hyperbolic growth. The method we present in this paper (HyperMap) does not use the Metropolis-Hastings algorithm or any heuristics to guide it, requires no manual intervention, it is simple, and it is based on a recent model of growing complex networks that has been shown to describe the evolution of different real networks well~\cite{PaBoKr11}.

A different mapping of the AS Internet to the hyperbolic plane was performed in~\cite{Shavitt2008}. The authors found that the hop lengths of the shortest AS paths in the Internet can be embedded into the hyperbolic plane with low distortion, and that the resulting embedding can be used for efficient overlay network construction and accurate path distance estimation. Our work is different from~\cite{Shavitt2008} in that hyperbolic distances between ASs in our case are not directly defined by their ``observable'' AS path lengths. Instead, they are defined by ``hidden'' popularity and similarity node coordinates that manifest themselves indirectly via the nodes' connections and disconnections. Section~\ref{sec:navigation} indicates that short paths follow well the underlying hyperbolic geodesics in our mapping. However, nodes at short AS path distances are not always hyperbolically closer than nodes separated by longer paths, and as we have seen in Section~\ref{sec:performance_comparison}, HyperMap performed quite differently from the Inverse Shortest Path (ISP) method.

While in this paper we have focused on the AS Internet, HyperMap may be applicable to other real networks (e.g., social networks) and to other interesting problems, such as the challenging problem of predicting future links in different evolving networks~\cite{LuZhou11}.  From a theoretical perspective, our results advance our understanding of mapping real growing networks to their hyperbolic spaces, a problem that so far has been solved only for static networks~\cite{BoPa10}.

\section*{Acknowledgments}
We thank M. Bogu{\~n}{\'a}, M. Kitsak and M.{\'A}. Serrano for many useful discussions, and B. Huffaker for help with the AS geographic data. This work was supported by a Marie Curie International Reintegration Grant within the 7th European Community Framework Programme, by an AWS in Education grant award, by NSF CNS-0964236 and CNS-1039646, by DARPA HR0011-12-1-0012, and by Cisco Systems.

\appendix

Here we consider a network that has grown up to $t$ nodes according to the E-PSO model and derive the expressions for: (i) the expected degree of node $i$ by time $t$, $\bar{k}_i(t)$ (Equation~\ref{eq:k_i_t}); (ii) the probability density of the node radial coordinate $f_t(r)$ (Equation~\ref{eq:f_r_i_t}); and (iii) the global connection probability  $\tilde{p}(x(t))$ (Equation~\ref{eq:global_prob}).

\underline{Expected degree of node $i$ by time $t$, $\bar{k}_i(t)$.} In both the basic and the generalized PSO models the expected degree of node $i$ by time $t$ satisfies $\bar{k}_i(t) \propto \left(\frac{i}{t}\right)^{-\beta}$, $0 < \beta \leq 1$, which means that the degree distribution is a power law $P(k)\propto k^{-\gamma}$, $\gamma=1+\frac{1}{\beta} \geq 2$~\cite{PaBoKr11}. We show below that the same result holds in the E-PSO model.

First recall from~\cite{PaBoKr11} that in the basic PSO model the probability that an existing node $i$ attracts a link from a new node $l >i$ is $\Pi(i,l)=m \frac{e^{-\frac{\zeta}{2}r_i(l)}}{\int_{1}^{l} e^{-\frac{\zeta}{2}r_i(l)}di}$, where $r_i(l)=\beta r_i+(1-\beta)r_l$, $r_i=\frac{2}{\zeta}\ln{i}$, $i \leq l$. In E-PSO, since new node $l$ brings in $\bar{m}_l(t)$ new links (Equation~(\ref{eq:m_i})) instead of $m$, this probability becomes:
\begin{eqnarray}
\label{eq:attraction_prob_external}
\nonumber \Pi(i,l,t)&=&\bar{m}_l(t) \frac{e^{-\frac{\zeta}{2}r_i(l)}}{\int_{1}^{l} e^{-\frac{\zeta}{2}r_i(l)}di}=\bar{m}_l(t) \frac{(\frac{i}{l})^{-\beta}}{l I_l}\\
\nonumber &=& \left[m+\bar{L}_l(t)\right]\frac{(\frac{i}{l})^{-\beta}}{l I_l} \approx \frac{m}{I_t} l^{\beta-1}i^{-\beta}\\
&+& \frac{2L}{I_t^2(2\beta-1)}\left[\left(\frac{t}{l}\right)^{2\beta-1}-1\right]l^{\beta-1}i^{-\beta}.~~
\end{eqnarray}
For the approximation above we used that for large $l, t$, $I_l=\frac{1}{1-\beta}(1-l^{-(1-\beta)}) \approx \frac{1}{1-\beta}(1-t^{-(1-\beta)})=I_t$.
Using Equation~(\ref{eq:attraction_prob_external}) and the fact that node $i$ brings in on average $\bar{m}_i(t)$ links when it first appears (Equation~(\ref{eq:m_i})), we can write: $\bar{k}_i(t)=\bar{m}_i(t)+\int_{i}^{t}\Pi(i,l,t)dl$, where
\begin{eqnarray}
\nonumber \int_{i}^{t}\Pi(i,l,t)dl \approx \frac{m}{I_t\beta}\left[\left(\frac{i}{t}\right)^{-\beta}-1\right]\\
\nonumber +\frac{2L}{I_t^2(2\beta-1)}\left[\frac{2\beta-1}{\beta(1-\beta)}\left(\frac{i}{t}\right)^{-\beta}-\frac{1}{1-\beta}\left(\frac{i}{t}\right)^{1-2\beta}+\frac{1}{\beta}\right].
\end{eqnarray}
Since $0 < \beta \leq 1$ we have that $\bar{k}_i(t) \propto \left(\frac{i}{t}\right)^{-\beta}=e^{-\frac{\zeta}{2}(r_i(t)-r_t)}$. As in the PSO models this means~\cite{PaBoKr11} that in E-PSO the degree distribution is power law $P(k)\propto k^{-\gamma}$ with $\gamma=1+\frac{1}{\beta} \geq 2$. Finally, the resulting average node degree in E-PSO is:
\begin{eqnarray}
\nonumber \bar{k}&=&\frac{2}{t}\int_{1}^{t}\bar{m}_i(t)di=\frac{2}{t}m(t-1)+\frac{2}{t}\int_{1}^{t}\bar{L}_i(t)di\\
\nonumber &\approx& 2m+\frac{2L}{(1-t^{-(1-\beta)})^2}\left[\frac{t^{-2(1-\beta)}}{2\beta-1}-\frac{2 t^{-(1-\beta)}}{\beta}+1\right]\\
&\approx& 2(m+L).
\end{eqnarray}
The approximations above hold for large $t$.

\underline{Probability density of the node radial coordinate, $f_t(r)$.} Let $r(t)$ be a random variable denoting the radial coordinate of a node at time $t$. We can write:
\begin{eqnarray}
\label{eq:cdf}
\nonumber P(r(t) \leq r) &=& \frac{\textnormal{All nodes $i \leq t$ such that $r_i(t) \leq r$}}{t}\\
\nonumber &=&\frac{\textnormal{All nodes $i \leq t$ such that $r_i \leq \frac{r}{\beta}-\frac{1-\beta}{\beta}r_t$}}{t}\\
&=& \frac{e^{\frac{\zeta}{2}(\frac{r}{\beta}-\frac{1-\beta}{\beta}r_t)}}{e^{\frac{\zeta}{2}r_t}}=e^{\frac{\zeta}{2\beta}(r-r_t)}.
\end{eqnarray}
The first equality in Equation (\ref{eq:cdf}) is the percentage of nodes whose radial coordinate is less than (or equal to) $r$, and the second equality uses the fact that $r_i(t)= \beta r_i+(1-\beta)r_t$. To ease analysis we treat $r(t)$ as a continuous random variable, in which case its probability density function $f_t(r)$ is found by differentiating Equation (\ref{eq:cdf}) with respect to $r$: $f_t(r) = \frac{\zeta}{2\beta} e^{\frac{\zeta}{2\beta}(r-r_t)}$.

\underline{Global connection probability, $\tilde{p}(x(t))$.} Recall from Section~\ref{sec:preliminaries} that $x_{ij}(t)\approx r_i(t)+r_j(t)+\frac{2}{\zeta}\ln{(\theta_{ij}/2)}$ is the approximate relation for the hyperbolic distance between two nodes $i, j$ at time $t$.  Given that the youngest of the two nodes is node $i$, and using the fact that  $r_i(t)=\beta r_i +(1-\beta)r_t$, $\forall i \leq t$, the hyperbolic distance between the two nodes when $i$ appeared is $x_{ij}(i)= x_{ij}(t)+(2\beta-2)r_t-(2\beta-2)r_i$.  Since this relation holds for any $i, j$ pair and depends only on the index $i$, we can drop the subscript and write:
\begin{eqnarray}
\label{eq:shift_1}
\nonumber x(i) &=& x(t)+(2\beta-2)r_t-(2\beta-2)r_i\\
&=& x(t)+\frac{2}{\zeta}\ln{\left(\frac{t}{i}\right)^{2\beta-2}},
\end{eqnarray}
where $x(i) \geq 0$ when $i\geq i_{min}=te^{-\frac{\zeta x(t)}{4(1-\beta)}}$.

Recall that $R_i=r_i-\frac{2}{\zeta}\ln\left[\frac{2T}{\sin{T\pi}}\frac{I_i}{\bar{m}_i(t)}\right]$, where $I_i=\frac{1}{1-\beta}(1-i^{-(1-\beta)})$ and $\bar{m}_i(t)$ given by Equation~(\ref{eq:m_i}). Using Equation~(\ref{eq:shift_1}), we can write: $x(i)-R_i = x(t)-R_t+\Delta_i(t)$, where $\Delta_i(t)=\frac{2}{\zeta}\ln{\left[\left(\frac{t}{i}\right)^{2\beta-1}\frac{m I_i}{\bar{m}_i(t) I_t}\right]}$.
Now, given the hyperbolic distance between two nodes at time $t$, $x(t)$, and knowing that the youngest of the two nodes appeared at time $i$, the probability that these two nodes are connected is the probability that they were connected at time $i$:
\begin{equation}
\label{eq:p_x_i_appendix}
p(x(i))=\frac{1}{1+e^{\frac{\zeta}{2T}(x(i)-R_i)}} = \frac{1}{1+e^{\frac{\zeta}{2T}(x(t)-R_t+\Delta_i(t))}}.
\end{equation}
Removing the condition on the index $i$ from the above relation we get the  global connection probability $\tilde{p}(x(t))$:
\begin{eqnarray}
\label{eq:p_global_appendix}
\nonumber \tilde{p}(x(t))&=&\frac{1}{t-i_{min}+1}\sum_{i=i_{min}}^{t}p(x(i))\\
\nonumber &=& \frac{1}{t-i_{min}+1} \sum_{i=i_{min}}^{t}\frac{1}{1+e^{\frac{\zeta}{2T}(x(t)-R_t+\Delta_i(t))}}.\\
\end{eqnarray}
Since time $i$ is discrete, and a connection can occur only when $i \geq 2$, in Equation~(\ref{eq:p_global_appendix}) we use $i_{min}=\max(2, \lceil te^{-\frac{\zeta x(t)}{4(1-\beta)}}\rceil)$.

Finally, using that $I_i \approx I_t$ for large $i$, $t$, we can approximate $\Delta_i(t)$ in Eq.~(\ref{eq:p_x_i_appendix}) by $\Delta_i(t) \approx -\frac{2}{\zeta}\ln{\left[A+(1-A)\left(\frac{i}{t}\right)^{2\beta-1}\right]}$, where $A=\frac{(\bar{k}-2m)(1-\beta)}{m(2\beta-1)(1-t^{-(1-\beta)})}$. By performing now the Taylor series expansion of Eq.~(\ref{eq:p_x_i_appendix}) around $t$, one can check that $p(x(i))$ can be well approximated by the first term of the expansion, i.e., the term $\frac{1}{1+e^{\frac{\zeta}{2T}(x(t)-R_t)}}$, for a wide range of parameter values $m \leq \frac{\bar{k}}{2}$ and $\beta=\frac{1}{\gamma-1}$ ($\gamma \geq 2$). This means that $\tilde{p}(x(t))$ in Equation~(\ref{eq:p_global_appendix}) can be well approximated by $\tilde{p}(x(t)) \approx \frac{1}{1+e^{\frac{\zeta}{2T}(x(t)-R_t)}}$. As terms other than the first in the Taylor series are multiplied by powers of $(1-A)$, this approximation holds best when $|1-A| \leq 1$, i.e., when  $\gamma \leq 3$ and $m \geq \frac{\bar{k}}{2}\left(\frac{\gamma-2}{1-(3-\gamma)t^{-\frac{\gamma-2}{\gamma-1}}}\right)$, which hold in the AS Internet.

\bibliographystyle{IEEEtran}
\bibliography{bib}

\end{document}